\documentclass[12pt,preprint]{aastex}

\usepackage{epsfig}

\shorttitle{Line Survey of CIT\,6}
\shortauthors{Zhang, Kwok, \& Dinh-V-Trung}

\begin{document}

\title{A Molecular Line Survey of the Highly Evolved Carbon Star CIT\,6}

\author{Yong Zhang \& Sun Kwok} 
\affil{Department of Physics, University of Hong Kong, Pokfulam Road, Hong Kong}
\email{zhangy96@hku.hk; sunkwok@hku.hk}

\author{Dinh-V-Trung\altaffilmark{1}} 
\affil{Institute of Astronomy and Astrophysics, Academia Sinica\\
P.O Box 23-141, Taipei 106, Taiwan}
\email{trung@asiaa.sinica.edu.tw}

\altaffiltext{1}{on leave from Institute of Physics and Electronics, Vietnamese Academy of Science and Technology, 10 DaoTan Street, BaDinh, Hanoi, Vietnam.}

\begin{abstract} 

We present a spectral line survey of the C-rich envelope \object{CIT\,6} in the 
$\lambda$ 2\,mm and 1.3\,mm bands carried out with the Arizona Radio Observatory (ARO) 12\,m telescope and 
the Heinrich Hertz Submillimeter Telescope (SMT).  The observations cover the frequency ranges
of 131--160\,GHz, 219--244\,GHz, and 252--268\,GHz with typical sensitivity limit of $T_R<10$\,mK.  A total of 74 individual emission features are detected, of which 69 are identified to arise from 21 molecular species and isotopologues, with 5 faint lines remaining unidentified. 
Two new molecules (C$_4$H and CH$_3$CN) and seven new isotopologues
(C$^{17}$O, $^{29}$SiC$_2$, $^{29}$SiO, $^{30}$SiO, $^{13}$CS, C$^{33}$S, and C$^{34}$S)
are detected in this object for the first time.
The column densities, excitation temperatures, and fractional 
abundances of the detected molecules are determined using rotation diagram analysis. Comparison of the spectra of \object{CIT\,6} to that of \object{IRC+10216} suggests that the spectral properties of \object{CIT\,6} are  generally consistent with those of \object{IRC+10216}.
For most of the molecular species, the intensity ratios of the lines detected
in the two objects are in good agreement with each other. Nevertheless,
there is evidence suggesting enhanced emission from CN and HC$_3$N and depleted emission from HCN, SiS, and C$_4$H in \object{CIT\,6}. Based on their
far-IR spectra, we find that \object{CIT\,6} probably has a lower
dust-to-molecular gas ratio than  \object{IRC+10216}.
To investigate the chemical evolution of evolved stars, we compare
the molecular abundances in the AGB envelopes \object{CIT\,6}
and \object{IRC+10216} and those in the bright proto-planetary nebula
\object{CRL\,618}. The implication on the circumstellar chemistry
is discussed.

\end{abstract}

\keywords{
ISM: molecules --- radio lines: stars --- line: identification ---
stars: AGB and post-AGB --- stars: circumstellar matter ---
stars: individual (CIT\,6) --- surveys}

\newpage

\section{Introduction}

The late stages of stellar evolution from the asymptotic giant branch (AGB) to planetary nebulae (PN) are now recognized as an active period of chemical synthesis of molecules.  The detection and analysis of millimeter wave molecular emission lines are fundamental to the understanding of the physical conditions and chemical processes leading to chemical synthesis.  Due to the rapid evolution of the star, the changing physical conditions, including  dust, stellar winds, shock waves, UV emission and X-rays from the central star, etc. play different roles in circumstellar chemistry.
This leads to corresponding different circumstellar chemical compositions in different evolutionary stages.  The envelopes around C-rich stars, with their enhanced carbon abundance, provide a perfect cradle for molecule formation.
Hitherto, more than 60 molecular species have been detected in C-star envelopes \citep{glassgold96,olo97,cernicharo00,ziu07}, most of which were discovered 
through their rotational lines at millimeter wavelengths.

Recent improvement in telescope design and receiver performance
enable us to detect new molecular emission with a higher sensitivity,
and thus with the possibility of shedding new light on circumstellar chemistry.
The most frequently investigated C-star envelope is \object{IRC+10216},
which is one of the richest molecular sources in the sky.
Several molecular line surveys have been presented for this
object \citep[see][and the references therein]{cernicharo00,he08},
which was found to harbor extremely abundant carbon chain and
metal-containing molecules. \object{IRC+10216} has been
frequently used  as a standard reference for the chemical compositions
of late-type stars. This inevitably invites the issue whether
\object{IRC+10216} is a chemically unique late-type star.
To settle this question, we require systematic surveys of molecular line 
emission from other C-star envelopes.
In the present study, we report a spectral-line survey of
the C-star envelope \object{CIT\,6} at millimeter wavelengths.
This allows us to compare the similarity and difference
in the chemical compositions between the two C-star envelopes.

\object{CIT\,6} (\object{RW\,LMi, GL\,1403, IRC+30219, IRAS\,10131+3049}) was first discovered during the Caltech 2-$\mu$m sky survey and was among the 14 very red infrared-bright optical-faint sources found \citep{ulrich66}.
\object{CIT\,6} is characterized by its very low color temperature, implying that the star is surrounded by a very thick dust  envelope and has been identified as a long-period variable with a period of about 628 days  \citep{alksnis95}.
From the period-luminosity relation, \citet{cohen96} estimated the distance of \object{CIT\,6} to be
$400\pm50$\,pc, which is slightly more distant than  \object{IRC+10216}, which has a distance of between $\sim120$\,pc \citep{groen98} and $\sim150$\,pc \citep{gue99}.
\object{CIT\,6}  is believed to be more evolved and has a lower mass loss rate
compared to \object{IRC+10216} \citep[see][e.g.]{fukasaku94}. 

The large polarization found in the visible and infrared wavelengths
implies that the distribution of circumstellar material around the star
is asymmetric \citep[][]{kruszewski68,dyck71}.
Multi-wavelength imaging observations have been performed to
study the structure of the nebula around \object{CIT\,6}.
The optical images obtained by the Hubble Space Telescope (HST)
and the near-IR images of \object{CIT\,6}
obtained by the Keck-I telescope have revealed a bipolar dust envelope
and an elongated component with time-variable asymmetry \citep{monnier00}.
Several scattering arcs were revealed
by the HST-NICMOS imaging polarimetry \citep{schmidt02}.
These arcs are nearly concentric and extend to large stellar radii.
Mid-IR images of \object{CIT\,6} were obtained by 
\citet{lagadec05} using the ESO 3.6-m telescope.
A cometary-like feature was revealed in their 9.7\,$\mu $m image.

There have been several observations of molecular lines in
\object{CIT\,6} at millimeter wavelength.
\citet{henkel85} reported observations of a few molecular lines 
in \object{CIT\,6} and \object{IRC+10216} between 18 and 150\,GHz.
They found that relative abundances of observed molecules in the
two sources have no significant differences. Using the
Nobeyama 45\,m radio telescope,
\citet{fukasaku94} observed a few transitions in the frequency ranges
between 39--47\,GHz and 85--91GHz in a sample of evolved stars
including \object{CIT\,6} and found that the abundance of HNC increases
with the evolutionary stage of the stars.  
\citet{bujarrabal94} presented observations of 10 molecular transitions
in C-rich and O-rich circumstellar envelopes including \object{CIT\,6}
with the IRAM 30\,m radio telescope at 1.3\,mm, 2\,mm, and 3\,mm  windows.
A recent molecular line survey was presented by \citet{woods03} using
the SEST 15\,m and Onsala 20\,m telescopes. They found that
\object{CIT\,6} stands out from the other C-rich envelopes due to its high CN/HCN ratio and low HNC/HCN ratio.  To date, the  molecular species positively detected in \object{CIT\,6}
at millimeter wavelengths include
CO, $^{13}$CO, CN, $^{13}$CN, CS, SiO, SiS, $^{29}$Si$^{32}$S, 
C$_{2}$H, SiC$_{2}$, HCN, H$^{13}$CN,
HNC, C$_{3}$N, HC$_{3}$N, HC$^{13}$CCN, HCC$^{13}$CN, and
HC$_{5}$N.

In this paper, we present the first systematical line survey
of \object{CIT\,6} at the 2\,mm and 1.3\,mm windows, using
the Arizona Radio Observatory (ARO) 12\,m telescope and
the Heinrich Hertz Submillimeter Telescope (SMT). The observations
are described in Sec.~2. In Sect.~3 we present the identifications 
and abundance calculations of the detected molecular species.
In Sect.~4 we discuss the implication of our findings on
circumstellar chemistry. The conclusions are given in Sect.~5.

\section{Observations and data reduction}

The observations were carried out in
beam switching mode with an azimuth beam throw of 2$'$
during the period from 2005 April to 2006 September. 
Pointing was checked by observations of a planet every 2 hours.
The  131--160\,GHz (2\,mm window)
spectra were obtained with the ARO 12\,m telescope at Kitt Peak,
and the 219--244\,GHz and 252--268\,GHz (1.3\,mm window) spectra
with the SMT 10\,m telescopes on Mount Graham, Arizona.
The 2- and 1.3-mm dual-channel SIS receivers were employed,
operated in single sideband dual polarization mode with a typical image 
rejection ratio of $>18$\,dB. 
At the ARO 12\,m, the spectrometer back-ends
were two 256-channel filter banks (FBs) with a channel
width of 1\,MHz
and a millimeter autocorrelator
(MAC) with 3072 channels and 195\,kHz per channel.
The spectrometers utilized at SMT were a 2048-channel
acousto-optical spectrometer (AOS) with a channel width of
500\,kHz and 1024-channel Forbes Filterbanks (FFBs)
with a channel width of 1\,MHz. 
The system noise temperatures were typically
150--400\,K at 2\,mm and 400--700\,K at 1.3\,mm.
The temperature scales at the ARO 12\,m and the SMT, $T^*_R$ and
$T^*_A$,  were obtained using standard vane calibration.
A $\sim15\%$ calibration error was obtained from
a few strong lines detected in different spectrometers and
different epochs.
The main beam brightness temperatures were derived through
$T_R=T^*_R/\eta^*_m$ and $T_R=T^*_A/\eta_{mb}$, for the
12\,m and the SMT data, respectively, where 
$\eta^*_m$ is the corrected beam efficiency ($\sim0.75$;
see the ARO 12\,m manual for its definition)
and $\eta_{mb}$ the beam efficiency ($\sim0.7$). Both telescopes have a large
beam size, covering the whole emission region.

The CLASS software package in GILDAS \footnote{GILDAS is developed and 
distributed by the Observatoire de Grenoble and IRAM.} was used to 
reduce the spectra. After discarding the bad scans  which are seriously affected by bandpass irregularities, we co-added the spectra from individual scans. The baseline was fitted with a low-order polynomial.  In order to improve the signal-to-noise ratios,
the spectra were smoothed and rebinned by a factor of 3,
 yielding a typical $rms$ noise temperature of
$<10$\,mK in main beam brightness temperature units and a spectral resolution of
$\sim 4$\,km/s. 
Since each line has been observed by two
different spectrometers, we can reduce the
uncertainties caused by ripples and bad channels.

\section{Results}

\subsection{Line identification and measurement}

A total of 74 individual features were detected in our survey, including some less than certain detections.  
Line identification is mainly based on the JPL catalog
\citep{pickeet98}\footnote{http://spec.jpl.nasa.gov.} and the
Cologne database for molecular spectroscopy
\citep[CDMS,][]{muller01,muller05}\footnote{http://www.ph1.uni-koeln.de/vorhersagen/.},
which give the molecular line frequencies based on theoretical calculations.
We also used the NIST Recommended Rest Frequencies for Observed Interstellar Molecular Microwave Transitions\footnote{http://physics.nist.gov/cgi-bin/micro/table5/start.pl}, which provides the molecular line frequencies
derived from observations of various sources.
Moreover, we utilized the recent observations of \object{IRC+10216} and 
\object{CRL\,618} by \citet{he08}, \citet{cernicharo00}, and \citet{pardo07} for reference. 
The complete spectra obtained with the
ARO 12\,m telescope and the  SMT 10\,m telescope are plotted in 
Figs.~\ref{spe_cit6_12m} and \ref{spe_cit6_smt} along with identified 
lines marked. Figs.~\ref{pro_12m} and \ref{pro_smt} give the 
detected molecular line profiles. Note that  the
two features at 221.459\,GHz and 241.458\,GHz clearly seen
in Fig.~\ref{spe_cit6_smt} are the strong CO lines from the image sideband.
Among the detected features,
69 are identified to arise from 21 molecular species and their isotopic 
variants.
Table~\ref{line} lists their assignments, peak and integrated intensities 
and line widths
species-by species. The half maximum line widths ($FWHM$) were 
obtained by fitting Gaussian line profiles.
If a line has an unresolved hyperfine structure, the peak intensity of the 
strongest transition is given, and the integrated intensity and
the $FWHM$ are those of the combined feature.

Five faint features remain unidentified and they are listed in Table~\ref{uline}.  Comparing with the unidentified line listing in the NIST frequency table, we find that the 148444, 255940, and 265936 MHz lines have no previous recorded detections elsewhere, and the 257035 and 262255 MHz may have corresponding U lines at 257015 and 262273 MHz detected in Sgr B2 (N) \citep{num98}.

A detailed description of the molecules detection is given below.

$CO$.
The $J=2$--1 transitions of CO, $^{13}$CO, and C$^{17}$O were detected 
at the 1.3\,mm window. 
 Fig.~\ref{pro_smt} clearly shows that the shape of the CO line
differs from that of the $^{13}$CO line. The CO line shows a
parabolic profile, while the  $^{13}$CO line has a double-peaked
profile, suggesting the former is optically thick, while the latter
is optically thin. \citet{teyssier06} observed the CO transitions
from $J=1$--0 to $J=6$--5 in \object{CIT\,6}. The shape of the
CO (2--1) line obtained by our observations is very similar to their
results (see their Fig.~6). The CO (2--1) and $^{13}$CO (2--1) 
lines have also been observed by \citet{groenewegen96} using
the IRAM telescope in 1991.  They obtained the integrated intensities
of the CO (2--1) and $^{13}$CO (2--1) lines to be 230 and 16.5\,K\,km\,s$^{-1}$ 
respectively,
a factor of $\sim3$ larger than our results.
As the beam size of the IRAM telescope is about half of that of 
the SMT, the discrepancy of the CO intensity can be explained
with different beam dilution effects. Hence, we can conclude that the
variation of CO intensity is relatively small
(within 10$\%$) during the past 15 years.
The extremely faint C$^{17}$O emission should be a new
detection for this object.

$SiC_2$.
SiC$_2$  has been detected by \citet{woods03} in the spectra
of \object{CIT\,6} through the 
SiC$_2$ (5$_{0,5}$--4$_{0,4}$)) transition at 115.382\,GHz.
In our line survey,
a number of SiC$_2$ transitions were prominently detected at the 2\,mm and 1.3\,mm windows.
These lines have a similar profile with an average linewidth of
$\Delta V_{FWHM}=23.9\pm1.9$\,km\,s$^{-1}$. We also detected two
faint $^{29}$SiC$_2$ transition although their intensities are only
at a 2--3$\sigma$ noise level. We cannot find  previous papers
reporting on the detections of these SiC$_2$
and $^{29}$SiC$_2$ transitions tabulated in Table~\ref{line}.

$CN$.
\object{CIT\,6} is characterized by bright CN emission.
The frequencies of three CN (2--1) fine-structure groups
lie in the region of the SMT spectrum. All of them were clearly seen
in Fig.~\ref{spe_cit6_smt}. The three CN (2--1) fine-structure groups
consist of 18 hyperfine structure components. 
\citet{bachiller97} carried out a survey of CN (2--1) and (1--0)
emission in a sample of evolved stars. They detected the strongest
two CN (2--1) groups in \object{CIT\,6}. Their observations suggests
the intensity ratio of high- and low-frequency fine-structure groups
to be 1.5, in excellent agreement with our result of 1.6.

$CS$.
Strong CS (3--2) emission was detected in the 2\,mm window.
This line had previously been discovered by \citet{bujarrabal94}
and \citet{henkel85}.
We also clearly detected its isotopic transitions, $^{13}$CS (3--2, 5--4) and
C$^{34}$S (3--2, 5--4). The faint  C$^{33}$S (5--4) emission was
only marginally discovered.
The CS (5--4) transition at 244.9\,GHz does not lie in the 
frequency region observed
in the work because of a gap between 244.5--252.5\,GHz in the survey.
Figs.~\ref{pro_12m} and \ref{pro_smt}  show some evidences that 
the line profiles of CS and its isotopologue emission are
slightly asymmetric in shape with a brighter red wing.

$SiS$.
The $J=6$--5 and $J=5$--4 transitions of SiS in \object{CIT\,6} have previously 
been observed by many researchers \citep{henkel85,bujarrabal94,woods03,schoier07}.
In this work, three  SiS  transitions with higher $J$ 
(8--7, 13--12, and 14--13) were 
clearly detected with an average linewidth of
$\Delta V_{FWHM}=22.6\pm2.2$\,km\,s$^{-1}$.

$SiO$.
SiO emission in this carbon star has been extensively studied. The 
transitions observed by previous studies include $J=2$--1, 3--2, 5--4, 
and 8--7 \citep{bujarrabal94,bieging00,woods03,schoier06}.       
In this work, we report the detection of the
SiO (6--5) transition and its $^{29}$S and $^{30}$S isotope lines
as well. The SiO (6--5) line detected in this survey has a similar width to 
other SiO transitions detected by previous researchers.

$C_2H$.
The $N=1$--0 transition of C$_2$H  in this object has been detected
by \citet{fukasaku94}.  The $N=3$--2 transition
lies in our surveyed frequency range. The rotation transition is
split in six hyperfine-structure lines grouped in three fine-structure
groups, all of which were detected at  the 2\,mm window.
The two main components (3$_{7/2}$--2$_{5/2}$ and 
3$_{5/2}$--2$_{3/2}$) are quite strong and have a well-defined profile,
which  is similar to that of the CS line with higher flux in the red wing.
The 3$_{7/2}$--2$_{5/2}$ transition is blended with a weak 
SiC$_2$ line.  The 3$_{5/2}$--2$_{5/2}$ transition is too faint to
obtain a reliable intensity.

$HCN$.
The $J=3$-2 transitions of HCN and H$^{13}$CN were clearly detected.
The HCN  (3--2) transition is the second brightest line after
the CO (3--2) transition in our survey. 
This line was first observed by \citet{bieging00} using the SMT, who obtained an integrated
intensity of 43.1\,K\,km\,s$^{-1}$, in excellent agreement with our
measurement. Given its line shape, the line should be optically thick.
There are five favorable vibrationally excited lines of HCN
present in the surveyed frequency range.
Three of them were clearly detected, as illustrated in Table~\ref{line}.
The weak $\nu_2=2^{2f},2^{2e}$ $J=3$--2 transitions are below the
3$\sigma$ noise level. In \object{CIT\,6}, HCN is the only
species with detected vibrationally excited transitions.
Fig.~\ref{pro_smt} shows that these
vibrationally excited lines are probably narrower than the HCN  (3--2) line.
Discarding the uncertain detection of the  $\nu_2=2^{2f},2^{2e}$ $J=3$--2 
transitions and the $\nu_2=1^{1e}$ $J=3$--2 transition which is partially
blended with the HCN (3--2) line, we obtain the average width of the 
vibrationally excited lines of HCN to be 14.2\,K\,km\,s$^{-1}$.
The narrow linewidth of the vibrationally excited lines
suggests that they might arise from the hot inner region with low expansion velocity.
We do not detect the vibrationally excited lines of H$^{13}$CN.

$C_3N$.
\citet{fukasaku94} and  \citet{woods03} observed
the C$_3$N (11--10 a,b) transitions in \object{CIT\,6}.
Both lines
are relatively weak. Due to fine-structure interactions,
 every rotational transition of C$_3$N is split into doublets
of similar intensity. 
There are 12 C$_3$N transitions present in our survey range with
six at the 2\,mm window and six at the 1.3\,mm window. We detected the
strongest four C$_3$N transitions (14--13 a,b and 15--14 a,b) at the 2\,mm 
window with $T_R>20$\,mK. The 16--15 a,b transitions at around 158\,G\,Hz
have a comparable intensity with the other four transitions at
the 2\,mm window. However, they fall within a spectral
region with a high noise level, and thus are not listed in
Table.~\ref{line}.

$C_4H$.
To our knowledge, C$_4$H has not been detected in \object{CIT\,6}
before this work. \citet{fukasaku94} and  \citet{woods03}
estimated the intensity upper-limits of the
C$_4$H (10--9 a,b) transitions, which indicate
that C$_4$H emission in this object is relatively weak.
Analogous to that of C$_3$N,
every rotational transition of C$_4$H is split into two components with
a similar intensity.
There are 12 favorable lines of C$_4$H present in the frequency range
surveyed here. Eight of them were detected with $T_R=9$--35\,mK.
The C$_4$H (15--14 a,b) and (16--15 a,b) transitions were clearly
detected with well-defined profiles at the 2\,mm window.
The C$_4$H (14--13 a,b) lines fall within a spectral region with a high
noise level, and thus are not listed in Table~\ref{line}.
The four C$_4$H lines (24--23 a,b and 28--27 a,b) at the 1.3\,mm window 
are extremely faint, and are only marginally detected. The
C$_4$H (25--24 a,b) transitions at around 238\,GHz are overwhelmed by
noise.

$HC_3N$. Seven HC$_3$N transitions from $J=15$--14 to $J=29$--28
are in the frequency range of our survey. All of them were prominently detected with 
$T_R>45$\,mK. Previously only lower $J$ transitions of  HC$_3$N have been reported
 \citep{henkel85,bujarrabal94,fukasaku94,woods03} and to the best of our knowledge all these detections are new.
Because of a gap in the 1.3\,mm spectrum, the 
HC$_3$N (27--26) at 245.606\,GHz was not detected here.
As shown in Table~\ref{line}, there is a trend that the widths of
HC$_3$N lines decrease with increasing $J$-values, suggesting that
the high-$J$ transitions might originate in hot inner regions 
which have a lower velocity compared to the regions from where
the low-$J$ transitions arise. All HC$_3$N lines detected
in this survey show asymmetric profiles (see Fig.~\ref{pro_12m} 
and \ref{pro_smt}). Opposite to CS and C$_2$H lines,
HC$_3$N lines have a brighter blue wing relative to the red one.
This indicates that the molecular envelope is asymmetric
along the line of sight, i.e., the chemical compositions and/or
physical conditions in the red and blue sides are different.

$CH_3CN$.
There are 14 favorable CH$_3$CN transitions in the survey
region. We detected the blended features of the (12$_1$--11$_1$)
and the (12$_0$--11$_0$) transitions with $T_R\sim23$\,mK, which are the 
strongest two among the 14 CH$_3$CN transitions.
 \citet{woods03} failed to detect CH$_3$CN in this object and
only estimated the intensity upper limit of the (6$_1$--5$_1$)
transition.  This is therefore the first detection of CH$_3$CN in \object{CIT\,6}.


All the lines discovered in this survey have been detected in
\object{IRC+10216} \citep{he08,cernicharo00}. Compared to
the spectrum of \object{IRC+10216} in the same frequency
range, the non-detected species in the spectrum of
\object{CIT\,6} include 
C$_3$H, C$_3$H$_2$, C$_2$S, C$_3$S, H$_2$CO, SiC, SiN, PN, and
metal containing molecules, all of which
have only weak emission in the spectra
of \object{IRC+10216}, and thus are below our detection limit.
A number of vibrationally excited species have been detected
in \object{IRC+10216} \citep[see][and the references therein]{cernicharo00}.
For \object{CIT\,6}, however, no vibrationally excited lines
except those of HCN are strong enough to be detected.
Moreover, we find no evidence for the presence of ionic species
in \object{CIT\,6}.

\subsection{Rotation diagram analysis and fractional abundances}

The standard ``rotation-diagram" method was applied to calculate
the excitation temperatures  ($T_{ex}$) 
and column densities ($N$) of the molecules observed
in our spectra. 
From the equation of  radiative transfer and assuming that
the lines are optically thin, the level populations are
in local thermal equilibrium (LTE), and $T_{ex}>>T_{bg}$, where $T_{bg}$ is the cosmic background
radiation temperature (2.7\,K), we have the well-known relation,
\begin{equation}
\ln \frac{N_u}{g_u}=\ln\frac{3k\int T_s dv}{8\pi^3\nu S\mu^2}=
\ln\frac{N}{Q(T_{ex})}-\frac{E_u}{kT_{ex}}.
\end{equation}
 $N_u$, $g_u$, and  $E_u$ are
the population, degeneracy, and excitation energy of the upper level,
$\int T_s dv$ is the integration  of the source brightness temperature
over the velocity, $S$ is the line strength,
$\mu$ is the dipole moment, $\nu$ is the line frequency, and
 $Q$ is the rotational partition function. 
If several transitions arising from levels covering a wide energy range
are observed, $T_{ex}$ and $N$ can be deduced using
a straight-line fit to ${N_u}/{g_u}$ versus ${E_u}/{kT_{ex}}$.

The rotation-diagram provides important tools for studies of
excitation conditions. 
Departure from the linear relation can be caused by different
excitation mechanisms or misidentification.
For SiC$_2$, SiS, HC$_3$N, and C$_4$H, there are adequate numbers of detected transitions covering a wide range of excitation energy and their rotation diagrams are given in Fig.~\ref{dia_cit6}.
Good linear correlations were obtained for these species. 
For the calculations, we have
corrected the effect of beam dilution through
 $T_s=T_R(\theta^2_b+\theta^2_s)/\theta^2_s$, where $\theta_b$ is
the antenna full beam at half-power ($\sim40${\arcsec} and 30{\arcsec}
for the ARO 12\,m and the SMT respectively) and $\theta_s$ is the source diameter.
$\theta_s$ may be different for different species. We
followed the assumption by \citet{fukasaku94} and took an common
$\theta_s$ of $20''$ for all the species.  Since the source size is likely to vary from species to species \citep{lin00}, this assumption will introduce a $\sim50\%$ uncertainty
in the derived column densities. 
The derived excitation temperatures and column densities are given in Table~\ref{col_cit6}. For these
species for which only one line was detected or observed
transitions arise from a narrow range of energy levels, 
the method of rotation diagram method cannot be employed and a constant $T_{ex}$
of 40\,K was assumed for the calculations of  their column densities.

Assuming that the molecular  envelope is a spherical shell,
the emission is optically thin, $T_{ex}$ is uniform 
throughout the envelope, mass loss rate and expansion velocity
are constant during the formation of the envelope, and
the molecular density follows an $r^{-2}$ law, we determined
the fractional abundances of the observed species with respect to
 H$_2$ through the formula proposed by \citet{olofsson96},
\begin{equation}\label{abundance}
f_{\rm X}=1.7\times10^{-28}\frac{v_e\theta_bD}{\dot{M}_{{\rm H}_2}}
\frac{Q(T_{ex})\nu_{ul}^2}{g_uA_{ul}}
\frac{e^{E_l/kT_{ex}}\int T_Rdv}{\int^{x_e}_{x_i}e^{-4\ln2x^2}dx},
\end{equation}
where $\int T_Rdv$ is given in K\,km\,s$^{-1}$,
the full half power beam width $\theta_b$ is in arc\,sec,
the expansion velocity $v_e$ is in km\,s$^{-1}$, $D$
is the distance in pc, $\dot{M}_{{\rm H}_2}$ is
the mass loss rate in M$_{\sun}\,{\rm yr}^{-1}$, $\nu_{ul}$  the
line frequency in GHz, $g_u$ is the statistical weight of
the upper level, $A_{ul}$ is the Einstein coefficient for the transition,
$E_l$ is the energy of the lower level, and $x_{i,e}=R_{i,e}/(\theta_bD)$
with $R_i$ and $R_e$ the inner radius and outer radius of the shell.
For the calculations, we adopted $D=400$\,pc \citep{cohen96}. 

For the determination of molecular abundances using eq.~2, we first derive the expansion velocity $v_e$=18 km s$^{-1}$ from the profile of the CO (2--1) line.  The mass loss rate of $\dot{M}_{{\rm H}_2}=3.2\times10^{-6}$\,M$_\sun ~{\rm yr}^{-1}$ is obtained by applying eq.~2 of \citet{winters02} to  the CO (2--1) line assuming $f_{\rm CO}=1\times10^{-3}$.  These values of $v_e$ and $\dot{M}_{{\rm H}_2}$ are in excellent agreement with those given by \citet{fukasaku94}.
These parameters  suggest that \object{CIT\,6} has a similar expansion velocity and about one order of magnitude lower mass loss rate in comparison with \object{IRC+10216}.

The resultant abundances are given in Table.~\ref{col_cit6}.
Combined with the uncertainties introduced by calibration, noise, baseline,
and source size, we estimate that the errors of the column densities and
abundances amount to a factor of $\sim 2$.
One should bear in mind that when the emission is optically thick
the $N$ and $f_{\rm X}$ listed in Table.~\ref{col_cit6} represent only lower limits.
A comparison of  our results with those derived by
\citet{fukasaku94} and \citet{woods03} are also given in  Table.~\ref{col_cit6} and no significant discrepancies are found.

\section{Discussion}

\subsection{Chemistry}
\subsubsection{Oxygen-bearing molecules}

Similar to \object{IRC+10216}, 
the molecular envelope of CIT 6 is also characterized by a lack of oxygen-bearing compounds and abundance of carbon-bearing compounds. In \object{IRC+10216},
three O-bearing molecules are observed,
CO, SiO and HCO$^+$ \citep{cernicharo00} whereas in \object{CIT\,6}, only two 
O-bearing molecules (CO and SiO) are detected.  We find that
the SiO (6--5)/$^{13}$CO (2--1) integrated intensity ratio in \object{CIT\,6} is 1.05,
in excellent agreement with the value of 1.13 in \object{IRC+10216} \citep{he08}.
In our discussion,  $^{13}$CO is taken as a reference molecule because
the $^{13}$CO (2--1) line is likely optically thin and the $^{12}$C/$^{13}$C isotopic
ratios are similar in \object{CIT\,6} and \object{IRC+10216} (see Sect.~4.2).
SiO has been commonly detected in C-rich
envelopes \citep{schoier06}, implying  
the presence of icy comets surrounding the stars
\citep{agundez06} and/or non-equilibrium chemical
processes \citep{cherchneff06}.
HCO$^+$ has a relatively low abundance
in \object{IRC+10216} \citep{glassgold96},
and is presumably below the current detection
limit even if it might be present in \object{CIT\,6}.

\subsubsection{Carbon-bearing molecules} \label{carbon}

We observed abundant carbon chains and
radicals in \object{CIT\,6}, including
CO, SiC$_2$, CN, HCN, CS, C$_2$H, C$_3$N, C$_4$H,
HC$_3$N, and CH$_3$CN, all of which are linear.
This characteristic feature is similar to those of  \object{IRC+10216} and
 \object{TMC~1} \citep[see][]{cernicharo00}
although these lines  are much fainter in \object{CIT\,6}.

The most intriguing characteristic of \object{CIT\,6}
is the strong CN emission. The integrated intensity ratio
of the CN (2--1) group and the $^{13}$CO (2--1) transition
is 4.6, a factor of 2.2 larger than the value in
\object{IRC+10216} \citep{he08}. CN is mainly formed
through the photodissociation of HCN,
\begin{equation}\label{hcn}
{\rm HCN+{\it h\nu} \rightarrow CN+H}.
\end{equation}
According to \citet{he08}, the H$^{13}$CN (3--2)/$^{13}$CO (2--1)
integrated intensity ratio in \object{IRC+10216} is 4.5,
a factor of 3.2 larger than that in \object{CIT\,6}.
Therefore, our observations provide strong evidence
that reaction~(\ref{hcn}) dominates the chemistry
of CN and HCN in AGB stars and the photodissociation
is more efficient in the more evolved C-rich envelope
\object{CIT\,6}.

The above discussion also suggests that about 30$\%$
CN formed from HCN has been destroyed. On the other hand,
CN can be reprocessed into HC$_3$N through the reaction
\begin{equation}
{\rm CN+C_2H_2 \rightarrow HC_3N+H}.
\end{equation}
We do find that 
the HC$_3$N line intensities relative to the
$^{13}$CO (2--1) transition  in \object{CIT\,6}
are a factor of $\sim3$ larger than those in
\object{IRC+10216}, indicating efficient formation of 
HC$_3$N in \object{CIT\,6}. 
We did not find evidence for the enhancement of the C$_3$N radical,
suggesting that photodissociation of HC$_3$N into C$_3$N  is
insignificant in this object.

\object{CIT\,6} shows strong C$_2$H emission. The
C$_2$H radical is dominantly produced through the 
photodissociation reaction
\begin{equation}\label{c2h2}
{\rm C_2H_2+{\it h\nu} \rightarrow C_2H+H}.
\end{equation}
Our observations show that the 
C$_2$H line intensities relative to the
$^{13}$CO (2--1) transition in \object{CIT\,6}
are almost the same at those in \object{IRC+10216},
suggesting that there is no significant C$_2$H enhancement 
in \object{CIT\,6} compared to \object{IRC+10216}.
This probably has an implication that C$_2$H is dominantly processed in 
early AGB stages. 
Since C$_2$H and HC$_3$N have the same
chemical precursor, the $f$(C$_2$H)/$f$(HC$_3$N) abundance
ratio can be used to test the chemical formation path
\citep{wootten80}. Our results yield a ratio of 4.2
for  $f$(C$_2$H)/$f$(HC$_3$N), which is in good agreement with
those found in interstellar clouds by \citet{wootten80}
and is consistent with the prediction of gas phase chemistry.

On the other hand, as shown in Figs.~\ref{pro_12m} and \ref{pro_smt},
HC$_3$N lines show profiles that differ from those of the C$_2$H lines.
This suggests that the $f$(C$_2$H)/$f$(HC$_3$N) ratio is
not a constant in the envelope. The chemistry structures
in the red and blue sides of the star are not identical.
This is probably a consequence of an inhomogeneous
density distribution or physical environment.
A complete understanding of the molecular environment
of \object{CIT\,6} calls for a comprehensive 3D photochemistry 
model and high-resolution mapping observations.

C$_4$H is positively detected in \object{CIT\,6}.
The C$_4$H radical can be formed via
\begin{equation}
{\rm C_2H_2+C_2 \rightarrow C_4H+H}.
\end{equation}
\citet{nejad87} suggested that ion-molecule reactions also
play an important role for the production of C$_4$H, unlike
those for HC$_3$N.
Based on millimeter interferometer observation,
\citet{dayal93} found that the photochemical model underestimates
the C$_4$H abundance in \object{IRC+10216} by a factor of 5.
Our observations suggest that \object{CIT\,6} has a $f$(C$_4$H)/$f$(HC$_3$N) 
abundance ratio of 3.1, which is lower than that in \object{IRC+10216}
\citep{he08} by a factor of $\sim3$. On the other hand, 
C$_4$H can be photodissociated into C$_2$H. However, as no
enhancement of  C$_2$H is found in \object{CIT\,6}, destruction
of C$_4$H should be insignificant in this C-rich envelope.
Therefore, the high radical abundance of C$_4$H in \object{IRC+10216}
still remains mystery.

CH$_3$CN is a new finding for this C-rich envelope. This molecule
is a symmetric top and has been widely used as diagnostics of
excitation temperature. We detected only one weak CH$_3$CN
line in \object{CIT\,6}, and thus cannot use it to derive an
excitation temperature.  CH$_3$CN can be produced via
\begin{equation}
{\rm CH^+_3+HCN \rightarrow CH_3CNH^+},
\end{equation}
followed by
\begin{equation}
{\rm CH_3CNH^++e \rightarrow CH_3CN+H}.
\end{equation}
The formation of CH$_3$CN may be very efficient in \object{CIT\,6}
since strong HCN emission is detected.
 \object{CIT\,6} has a
 CH$_3$CN (12--11)/$^{13}$CO (2--1) integrated intensity
ratio of 0.058, in good agreement with the value of 0.043
in \object{IRC+10216} \citep{he08}.

Strong CS emission has been detected in our survey. According to
\citet{willacy98}, CS can be rapidly destroyed by shocks
which might occur after a star leaves the AGB stage
and ejects material in a very fast wind
\citep{herpin02}. Therefore, the high abundance of CS in
\object{CIT\,6} suggests that shocks are not
important for the chemistry in this C-rich envelope.

\subsubsection{Silicon-bearing molecules}

Three refractory Si-bearing species (SiO, SiS, and SiC$_2$) were detected in
\object{CIT\,6}.  Although other Si-bearing species were detected in 
\object{IRC+10216} \citep{cernicharo00}, emission from
the other Si-bearing species is relatively faint and
should be below the detection limit of our observations of \object{CIT\,6}.

We find that the abundances of SiO and SiC$_2$ in
\object{CIT\,6} are similar to those in \object{IRC+10216} 
determined by \citet{he08}. Although the situation may be complicated by 
optical depth effects, the HCN/SiO line intensity ratio has the potential 
to provide a useful tool to discriminate between C-rich
and O-rich envelopes and is a good tracer of mass
loss rate for M and S stars \citep[see e.g.][]{bieging00}. The
HCN (3--2)/SiO (6--5) intensity ratio in \object{CIT\,6}
is 8.4, in good agreement the value of 9.7 in 
\object{IRC+10216} \citep{he08}. There is no evidence showing that
the HCN/SiO line intensity ratio has dependance on the 
mass loss rate of C-rich stars.
\citet{gonzalez03} and \citet{schoier06} found a
correlation between the mass loss rate and the SiO
abundance for AGB stars. This is described as freeze-out
of SiO molecules onto dust grains. The similarity of the 
SiO abundances in \object{CIT\,6} and \object{IRC+10216}
suggests that the depletion of SiO onto dust grains
might be insignificant for the two C-rich envelopes.

Our observations show that the SiS abundance in \object{CIT\,6}
is lower than that in \object{IRC+10216}. The 
SiS (14--13)/SiO (6--5) intensity ratio in \object{CIT\,6}
is 1.1, about half of that in \object{IRC+10216} 
found by \citet{he08}. \citet{schoier07} did not find a strong 
correlation between the mass loss rate and the SiS abundance,
suggesting that SiS molecules are less likely to be depleted onto dust grains than SiO molecules. Therefore, freeze-out should not be the reason for the depletion of
SiS in \object{CIT\,6} since no depletion of SiO is found for this object. On the other hand, with a
high efficiency, SiS can be photodissociated into Si$^+$, which
initiates circumstellar SiC$_n$ chemistry \citep{mackay99}.
Hence, we infer that efficient photodissociation in \object{CIT\,6}
has destroyed SiS, and silicon chemistry has been ongoing in this evolved C-star envelope.
Interferometric observations of these Si-bearing molecules
are obviously needed to verify the conjecture.

\subsection{Isotopic ratios}

Isotopic ratios of various elements provide substantial tests for
nucleosynthesis  of low- and intermediate-mass stars (LIMS). When
a LIMS evolves into the AGB stage, the nucleosynthesized
products synthesized through the CNO cycle inside the star are 
dredged up to the surface and then are ejected into
the circumstellar envelope. Consequently,
the isotopic composition in the circumstellar shell can be markedly
changed. Based on the fractional abundances proposed in 
Table~\ref{col_cit6}, we deduce the isotopic ratios 
(or their lower limits) of carbon, oxygen, silicon, and sulfur in 
\object{CIT\,6}.
The results are listed in Table~\ref{isoto_cit6}. The errors
estimated from the measurement and calibration are given.
For comparison,
we also list the isotopic ratios for \object{IRC+10216}
\citep{cernicharo00} and  the Sun \citep{lodders}.

\subsubsection{Carbon} 

The $^{12}$C/$^{13}$C abundance ratio is the most studied
isotopic abundance in LIMS. Standard stellar models predict
that the $^{12}$C/$^{13}$C abundance ratio can be significantly
increased during the nucleosynthesis and dredge-up processes
in the AGB stage. However, extensive observations have
shown that the $^{12}$C/$^{13}$C abundance ratios in
LIMSs are considerably lower than those expected by standard stellar models
\citep[e.g.][]{charbonnel98}.  \citet{charbonnel95} proposed
an extra mixing process to account for the low  $^{12}$C/$^{13}$C.
In low-mass AGB stars, the nonstandard mixing called cool bottom processing 
may decrease the  $^{12}$C/$^{13}$C ratio to $\sim4$ 
\citep{sackmann99,boothroyd99}. For AGB stars more massive than
$\sim4 M_\sun$, the hot bottom burning  may take place 
 and induce $^{12}$C/$^{13}$C to further decrease to its
equilibrium value of $\sim3.5$ \citep{frost98}.  Current
observations of the CO isotopologues in PNs \citep{balser02,josselin03}
suggest that
the $^{12}$C/$^{13}$C ratio is in the range of 2.2--40,
supporting the theory including nonstandard mixing processes.

Three $^{13}$C-bearing species have been detected in this survey,
including $^{13}$CO, $^{13}$CS, and H$^{13}$CN. However, their
main lines are likely optically thick. Therefore, the abundance
ratios of CO, CS, and HCN and their isotopologues only provide 
lower limits of the $^{12}$C/$^{13}$C ratio. Our results
are in good agreement with those presented by
\citet{sopka89} who derived the abundance ratios
$^{12}$CO/$^{13}$CO$\ga25\pm10$ and
H$^{12}$CN/H$^{13}$CN$\ga5.4$ for \object{CIT\,6}.

We have detected the rare isotopes, $^{12}$C$^{34}$S and 
$^{13}$C$^{32}$S. If the $^{32}$S/$^{34}$S abundance
ratio were known, we could obtain the $^{12}$C/$^{13}$C ratio
using the two optically thin species. \citet{cernicharo00}
found that the sulfur isotopic ratios in \object{IRC+10216}
are close to solar. Therefore, we reasonably assume that
the  $^{32}$S/$^{34}$S ratio in \object{CIT\,6}
is the solar value. It follows that we obtained the 
$^{12}$C/$^{13}$C ratio of $45.4\pm4.9$ in \object{CIT\,6},
which is in perfect agreement with that found in
\object{IRC+10216} and is significantly lower than the
solar value.
The $^{12}$C/$^{13}$C ratios found in these C-rich envelopes
are also lower than the value of $75\pm9$ in the Orion Bar
proposed by \citet{keene98} using the 
C$^{18}$O/$^{13}$C$^{18}$O abundance ratio. This is consistent
with the hypothesis that nonstandard mixing processes have
decreased the $^{12}$C/$^{13}$C ratios in the 
envelopes around AGB stars.

As shown in  Table~\ref{isoto_cit6}, different $^{12}$C/$^{13}$C values
are obtained for \object{CIT\,6} if different species are used for
the calculations.
If completely ascribing the $^{12}$C/$^{13}$C discrepancies 
found for \object{CIT\,6} to the opacity effects of the main lines,
we can estimate the optical depths of the
CO (2--1), CS (3--2), and HCN (3--2) lines, which are
1.3, 1.2, and 2.0, respectively.
If the isotopic lines are also optically thick, the optical
depths obtained here should be considered as lower limits.

\subsubsection{Oxygen}

The nucleosynthesis and dredge-up processes in the AGB stage
can lead to strong enrichment of $^{17}$O relative to
$^{16}$O and $^{18}$O
\citep[see][for a recent review]{busso06}. \citet{wannier87}
found that the $^{17}$O/$^{18}$O ratios in C-rich envelopes
are markedly higher than the terrestrial and interstellar values, but the 
$^{16}$O/$^{18}$O ratios are comparable to the solar value.

We have detected C$^{17}$O,
allowing us to derive the $^{16}$O/$^{17}$O ratio in \object{CIT\,6}.
The C$^{16}$O/C$^{17}$O abundance ratio give a lower limit of
$237\pm26$. We may also use the optically thin species
$^{13}$C$^{16}$O and $^{12}$C$^{17}$O
to derive the  $^{16}$O/$^{17}$O ratio. 
Assuming $^{12}$C/$^{13}$C$=45.4$ (see above), we obtain a
$^{16}$O/$^{17}$O ratio of $890\pm97$. \citet{cernicharo00}
did not obtain the oxygen isotopic ratios. 
\citet{kahane92} calculated the $^{16}$O/$^{17}$O ratio for a sample
of C-rich envelopes. They found that $^{16}$O/$^{17}$O$=840^{+450}_{-270}$
and $840^{+230}_{-170}$ for  \object{CIT\,6} and \object{IRC+10216},
respectively. These values are in good agreement with our result,
agree with each other, and  are lower
than the solar value by a factor of about three, which
is consistent with predictions of stellar models.

No $^{18}$O-bearing species was detected. 
The non-detection of the C$^{18}$O (2--1) transition at 219560\,MHz 
seems to suggest that $^{17}$O/$^{18}$O$>1$ in \object{CIT\,6}. However,
this result should be taken with some caution
since the C$^{18}$O line is very close to the edge of the spectrum.

\subsubsection{Silicon and sulfer}

The nucleosynthesis in LIMS is expected to hardly affect
the elements in the 3rd row of the periodic table,
such as Si and S. \citet{cernicharo00} indeed found that
the Si and S isotopic ratios in \object{IRC+10216} are
compatible with the solar values.

The silicon isotopes, $^{29}$Si and $^{30}$Si, have been
detected through faint emission from $^{29}$SiO, $^{30}$SiO, and 
$^{29}$SiC$_2$. Since the SiO and SiC$_2$ lines are probably
optically thick, the lower limits of the $^{28}$Si/$^{30}$Si
and $^{28}$Si/$^{29}$Si ratios were derived. 
Given the low abundances of $^{29}$Si and $^{30}$Si,
the $^{29}$Si/$^{30}$Si ratio is not affected by opacity effects.
We obtained $^{29}$Si/$^{30}$Si$=1.0\pm0.4$, comparable with
the values in \object{IRC+10216} and in the Sun.

The C$^{32}$S/C$^{34}$S ratio gives a lower limit of
$^{32}$S/$^{34}$S in \object{CIT\,6}.  C$^{33}$S was only
marginally detected. The two optically thin species,
C$^{33}$S and C$^{34}$S, give a $^{33}$S/$^{34}$S ratio
of $0.2\pm0.2$. Consequently, we did
not find a significant deviation of the S isotopic
ratios in \object{CIT\,6} from those in
\object{IRC+10216} and in the Sun.

\subsection{Is IRC+10216 unique?}

Given its brightness and abundant molecular emission, \object{IRC+10216}
is the most surveyed object and has frequently served as a standard
reference for studying circumstellar chemistry. A commonly asked question is whether  \object{IRC+10216} can  truly represent C-rich AGB stars. To investigate the problem, we
systematically compare the spectra of \object{CIT\,6} with
those of \object{IRC+10216} obtained in the same
survey program \citep{he08}.

We find that all lines detected in this work have also been observed in  
\object{IRC+10216},
and all the strong lines detected in  \object{IRC+10216} are seen in \object{CIT\,6}.  In 
Fig.~\ref{linecomp}, we compare the intensity ratios of the lines detected
in both objects. The results show that
the line intensity ratios in the two objects
are in agreement within one order of magnitude
with an average value of 0.08 with a standard deviation of 0.05.
The  intensity ratios, however, might be affected by beam dilution.
\object{CIT\,6} is a more compact object and thus suffers from a larger 
beam dilution effect. If we correct for the beam dilution effect,
the $I$(CIT\,6)/$I$(IRC+10216) ratios should increase by a factor
of $\xi$, where 
\begin{equation}
\xi=(1+\frac{\theta_b^2}{\theta_{\rm CIT\,6}^2})/(1+\frac{\theta_b^2}{\theta_{\rm IRC+10216}^2}).
\end{equation}
Assuming $\theta_{\rm CIT\,6}=20''$ and $\theta_{\rm IRC+10216}=30''$
\citep{fukasaku94} and using $\theta_b=40''$ and 30$''$ for the
ARO 12m and the SMT respectively, we obtain the $\xi$ values of 1.8 
and 1.6 for the 12\,m and SMT data, respectively. 
While recognizing that eq.~9 assumes uniform brightness temperature which is probably not realistic, the result does suggest that there is no large correction factor difference between the $\lambda$ 2 mm and 3 mm bands.
Our conclusion that for most of the species the $I$(CIT\,6)/$I$(IRC+10216) line ratios
are in good agreement therefore stands.
Consequently, we conclude that \object{IRC+10216} is indeed likely
representative of C-rich envelopes although comparisons with a larger
sample of objects will be needed to make a stronger statement. The richness of 
molecular lines in the spectra of \object{IRC+10216} is mainly due to its relatively nearby distance, and not due to any special circumstances.

Fig.~\ref{linecomp} also shows that there are a few molecular species 
for which the
$I$(CIT\,6)/$I$(IRC+10216) ratios depart from the average 
value. For the CN and HC$_3$N lines, the ratios are higher, whereas
for HCN, SiS, and C$_4$H, the ratios are lower. As discussed in 
Sect.~\ref{carbon}, this partly reflects the chemical evolution in
the circumstellar envelope around the more evolved AGB star 
\object{CIT\,6}. However, the cause of the abnormally strong
C$_4$H emission in \object{IRC+10216} remains unknown.

Far-IR spectra of \object{CIT\,6} and \object{IRC+10216} have 
been obtained by the ISO Long Wavelength Spectrometer (LWS)
\citep{schoier02,cernicharo96}. The rotational transitions revealed
by the ISO spectra can trace the inner regions of the 
circumstellar envelopes \citep{herpin02}.
In  Fig~\ref{iso_con},
we compare the far-IR spectra of the two objects. 
The spectra were retrieved from the ISO archive.
Inspection of
the figure shows that their far-IR spectra are dominated by
 thermal continuum emission from the dust with some superimposed molecular
lines.  \citet{lin00} fitted the dust continuum of IRC+10216 with a single blackbody of 510 K and the continuum of CIT 6 with two blackbodies of 1000 K and 510 K.  In the long wavelength region, the dust temperatures of the two objects are therefore almost identical.

Fig.~\ref{isocomp} gives the continuum-subtracted  ISO LWS spectra of the two C-rich envelopes. A number of lines from CO, HCN, H$^{13}$CN, and vibrationally excited HCN have been identified by \citet{cernicharo96} in the spectrum of \object{IRC+10216}. Fig.~\ref{isocomp} shows that
the relative flux ratios of the lines detected in
the two object are in good agreement
although the detection of some weak lines is difficult in  \object{CIT\,6} as it is fainter.
The far-IR line fluxes detected in \object{IRC+10216} are
a factor of about 10 higher than those in \object{CIT\,6},
consistent with that found for the molecular lines at millimeter 
wavelengths. However, we find that the far-IR dust continuum emission in 
\object{IRC+10216} is a factor of about 20 times higher than that of 
\object{CIT\,6}, as shown in Fig~\ref{iso_con}. This  probably 
suggests that \object{IRC+10216} has a larger 
dust-to-molecular gas ratio.

\subsection{Abundance variations expected in later evolutionary stages}

Since \object{CIT\,6} has been proposed to be a highly evolved AGB star on the verge to become a Proto-PN (or PPN) \citep{schmidt02}, it would be useful to compare the molecular abundances of \object{CIT\,6} with the corresponding molecular abundances in the evolved AGB star \object{IRC+10216} and a PPN.
The archetypical PPN \object{CRL\,618} 
has abundant molecular emission. It is one of the brightest molecular sources
in the sky and is an ideal target for investigating circumstellar
chemistry \citep[see e.g.][]{pardo07}.  All three objects are carbon rich, and are commonly assumed to belong to a common evolutionary sequence.
Any systematic difference in molecular abundances in these three objects can be used to infer on the processes of chemical synthesis and destruction.

Fig.~\ref{compare} gives the fractional abundances
of the species in the three objects
as a function of their photodissociation rates
taken from the UMIST database for a temperature
of 300\,K.
The abundances in \object{IRC+10216} and \object{CRL\,618}
 are taken from \citet{woods03} and \citet{pardo07}, respectively.  
To facilitate the comparison, we follow
\citet{pardo07} and give the molecular abundances
relative to HC$_3$N.  According to this figure, except for CN and HCN, 
the molecular abundances in \object{CIT\,6} are obviously closer to those in
\object{IRC+10216} than to \object{CRL\,618}. 
\object{CRL\,618} has lower abundances of SiO and
CS, suggesting that destruction by shocks and depletion onto
 dust grains may play an important role for the chemistry
in this object. Compared to the other two objects,
\object{CRL\,618}, having a B0 central star, is exposed in a stronger UV radiation field. However,
we do not find a correlation between the abundance differences of
these objects and the molecular photodissociation rates, suggesting that
the destruction by photodissociation is not a factor affecting the
chemistry in these objects during the AGB-PPN transition.
However, we do find that some molecules have been efficiently
reprocessed during this transition.
Since the dynamical timescale of PPN is $\sim$10$^3$ years, our results therefore suggest a
rapid change of the chemical compositions after the star evolves into the PPN stage.

We should note that this study is limited to simple molecules in the gas phase.  There is strong evidence from infrared spectroscopy that the solid-state phase chemistry is very active in the AGB-PPN evolutionary transition, with many aromatic and aliphatic compounds being formed in the circumstellar envelope \citep{kwok04}.  Even some simple gas-phase molecules (e.g., acetylene and benzene) are difficult to detect via rotational transitions in the mm/submm region, and complex organics even more so.  Consequently, the technique of rotational mm/submm spectroscopy is unable to serve as a complete diagnostic tool for circumstellar chemistry.  It is more useful as a probe of photo- or shock-chemistry in the PN phase, or during the formative stage of early AGB evolution.

\section{Conclusions}

The presence of rich molecular species around evolved stars provides an opportunity to
study the evolution of chemistry in circumstellar envelopes, which have been widely suggested as one of the main sources of organic compounds in space.  As part of our project of investigating
circumstellar chemistry, this paper reports a spectral line survey
of the carbon-rich envelope \object{CIT\,6}, covering the
frequency range between 131--160, 219--244, and 252--268\,GHz
with a high sensitivity.  A total of 74 lines are reported in the survey.  We identify 69 lines belonging to 21 different molecular species and isotopologues, most of which are carbon-bearing species. 
The new species include two carbon-chain molecules, C$_4$H and CH$_3$CN,
and seven C, O, S, and Si isotopologues. Several new transitions from known
species have been detected for the first time in this object.
 The species with the largest number of detected emission lines in our survey is SiC$_2$, which has 19 lines. It is followed by HC$_3$N, with 7 lines.

We find that the line profiles for some molecular species have different shapes, suggesting that the
chemical structure is asymmetric in the envelope.  A 
comprehensive 3D photochemistry model is required to account for
the line intensities and profiles in \object{CIT\,6}.

The excitation temperatures, column densities and abundances of the detected molecules are determined through rotation-diagram analysis. The spectra of \object{CIT\,6} are
characterized by a large CN/HCN abundance ratio. Our results suggest that there is evidence for the photodissociation of HCN and SiS and the formation of CN and HC$_3$N in the evolved AGB envelope. The strong SiO and CS emission may suggest that depletion onto dust grains
and destruction by shocks are insignificant in this object.
An abundance comparison with the PPN \object{CRL\,618} implies to a rapid chemical evolution after a star leaves the AGB stage.

In order to investigate whether the molecular environment of  \object{IRC+10216} is intrinsically unique,
we systematically compare its spectra with those of
\object{CIT\,6}. According to the comparison, we find that  
the molecular species can be classified into three groups,
a) for most of the species, the intensity ratios of individual
lines in the two objects are in good agreement with each other;
b) the emission from HC$_3$N and CN may be enhanced in
\object{CIT\,6};
c) the emission from SiS, HCN, and C$_4$H may be depleted
in \object{CIT\,6}. The differences of the line-intensity ratios
in the two objects are probably a consequence
of chemical evolution with the exception  of C$_4$H, for which
a high abundance in \object{IRC+10216} cannot be explained by
photochemical models. The ISO LWS spectra show that
\object{CIT\,6} has a lower continuum-to-line ratio
than  \object{IRC+10216}, suggesting that the latter might
have a larger dust-to-molecular gas ratio.

Using the same telescope settings, we also obtained the spectra of the 
AGB stars \object{IRC+10216} and
\object{CRL\,3068}, the PPN \object{CRL\,2688}, and the
young PN \object{NGC\,7207}.  A detailed study of the chemical
compositions in different evolutionary stages
will be published in a separate paper.

\acknowledgments

We are grateful to the ARO staff for their help during the observing run.
Otto Peng and Yu-Chin Huang assisted in the observations.  We also thank Jun-ichi Nakashima  and Jin-Hua He for useful discussions. We acknowledge an
anonymous referee for comments which helped strengthen the paper.
The work was supported by a grant from the Research Grants Council of the Hong Kong Special Administrative Region, China (Project No. HKU 7028/07P).

\begin{figure*}
\epsfig{file=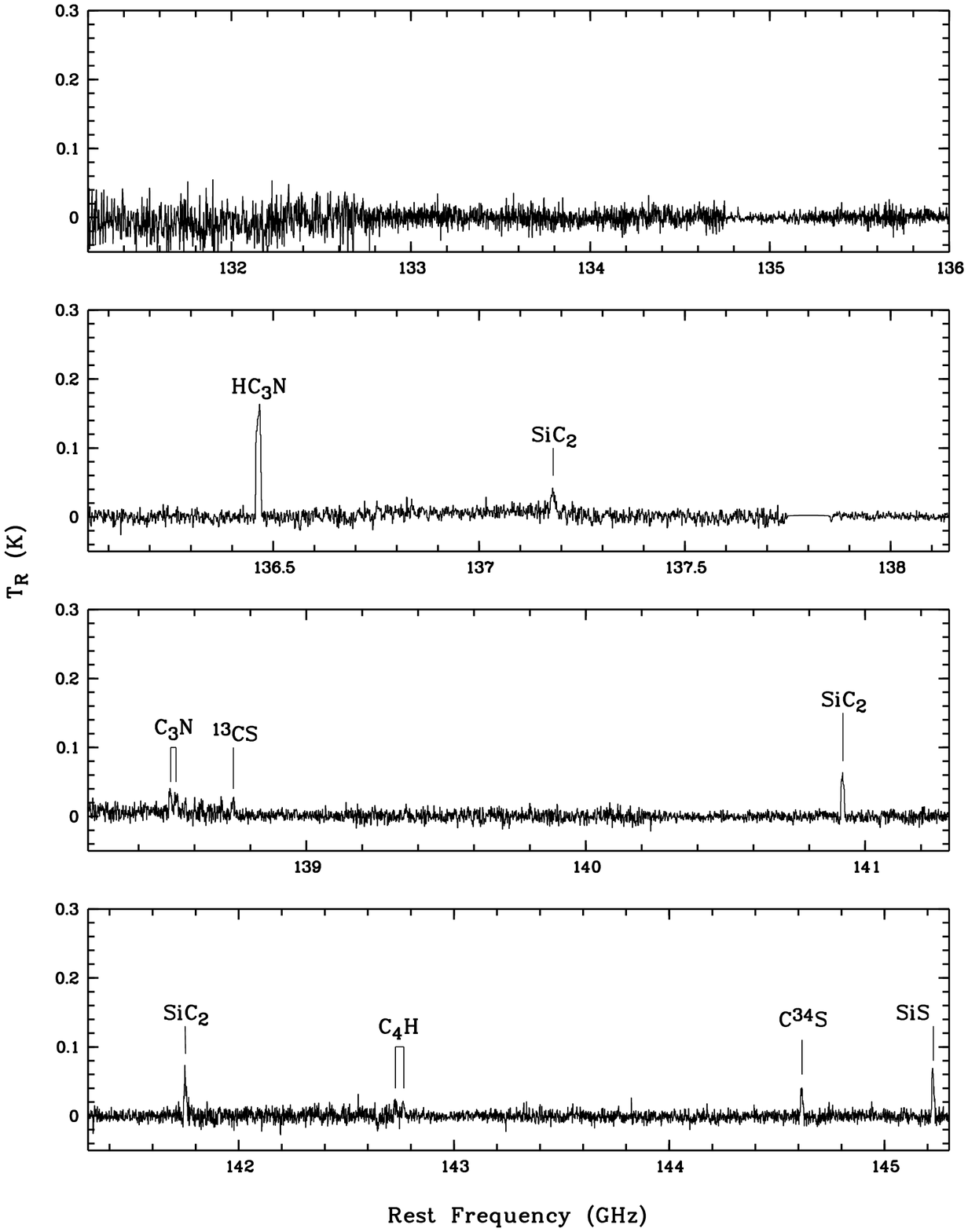,
height=20cm, }
\caption{Spectrum of CIT\,6 in the frequency range
131--160\,GHz obtained with the ARO 12\,m telescope. 
The spectra have been smoothed to a resolution of 1\,MHz.
}
\label{spe_cit6_12m}
\end{figure*}

\addtocounter{figure}{-1}
\begin{figure*}
\epsfig{file=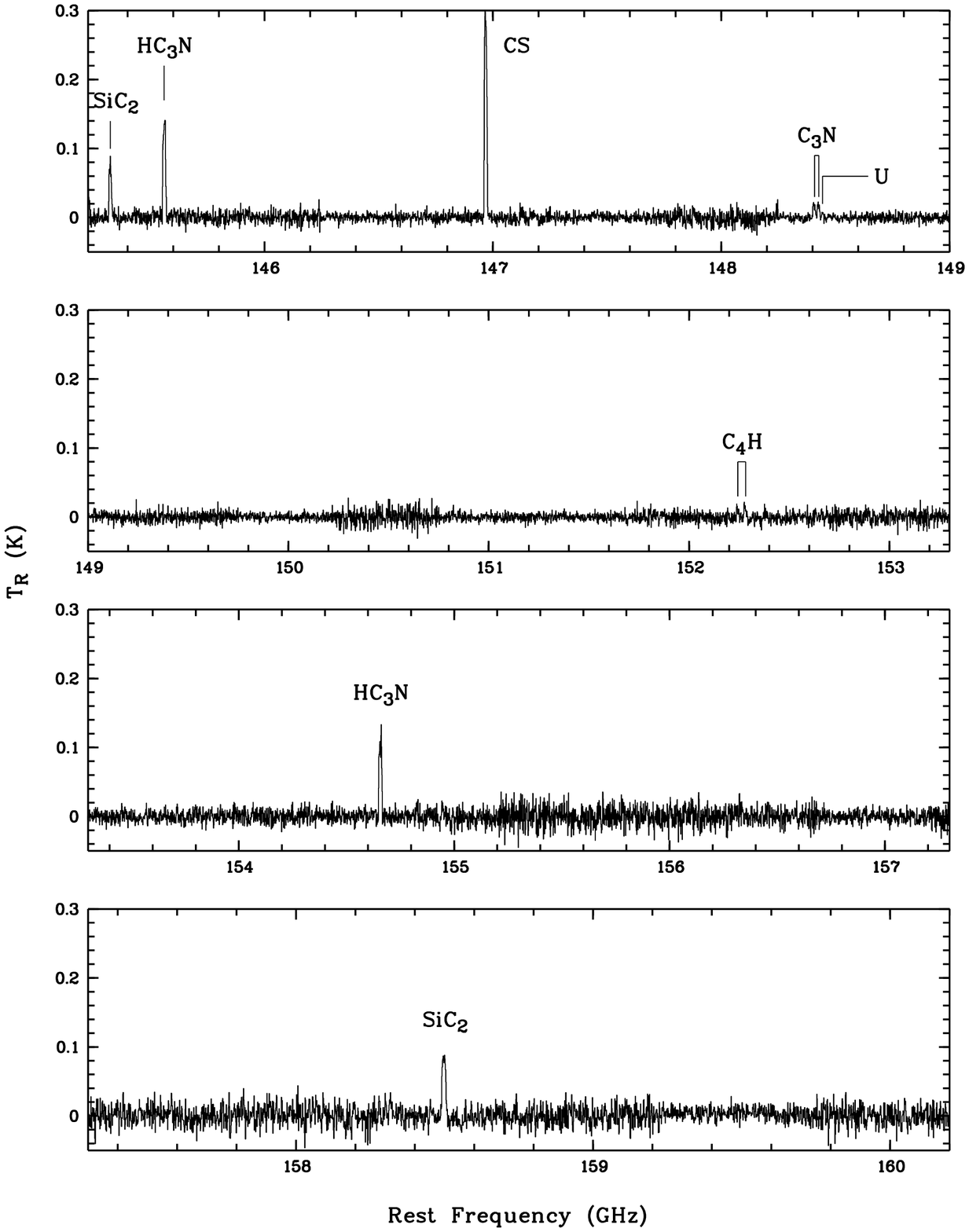,
height=20cm, }
\caption{continued.}
\end{figure*}

\begin{figure*}
\epsfig{file=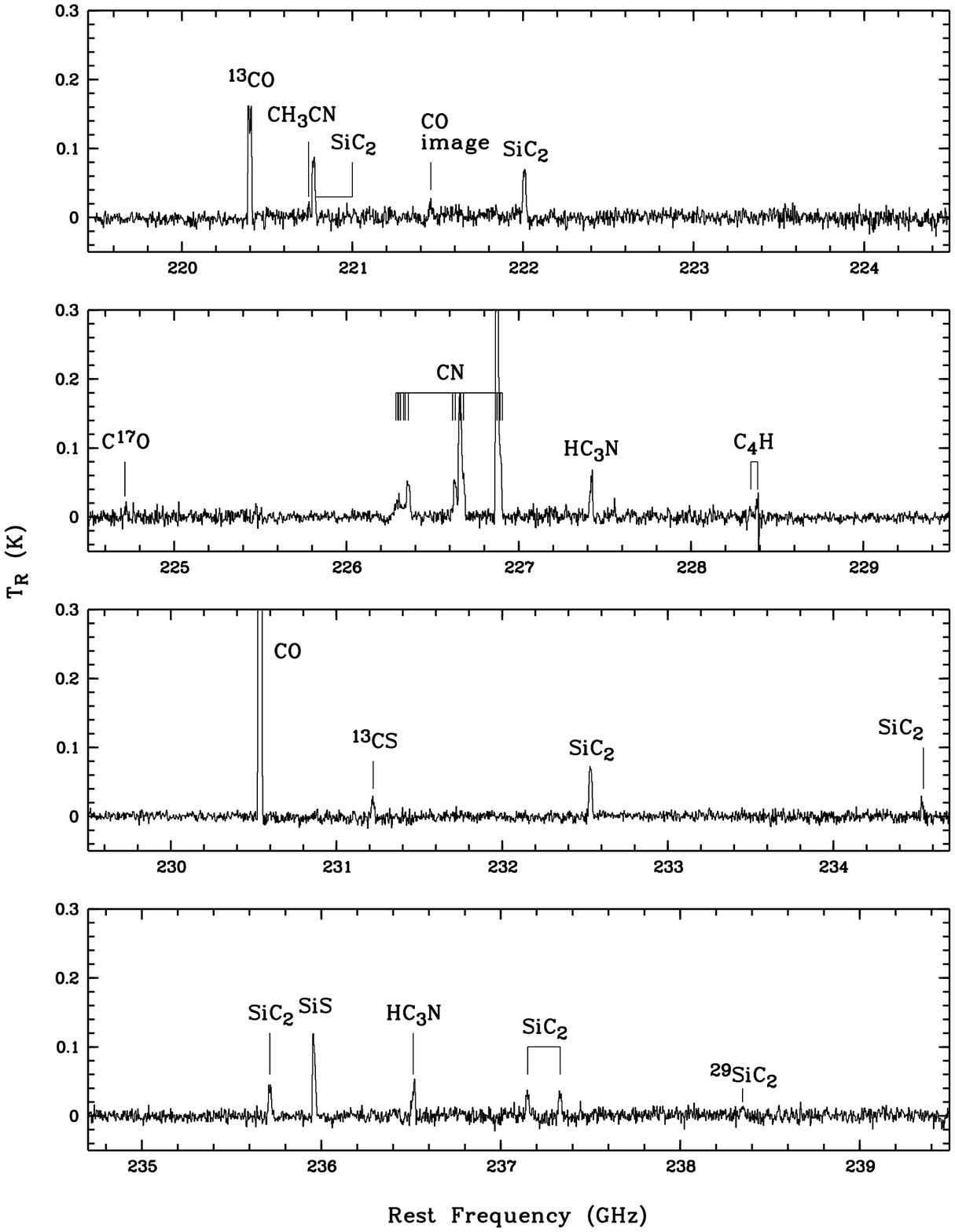,
height=20cm, }
\caption{Spectrum of CIT\,6 in the frequency ranges
219--244\,GHz and 252--268\,GHz 
obtained with the SMT 10\,m telescope.  
The spectra have been smoothed to a resolution of 3\,MHz.
}
\label{spe_cit6_smt}
\end{figure*}

\addtocounter{figure}{-1}
\begin{figure*}
\epsfig{file=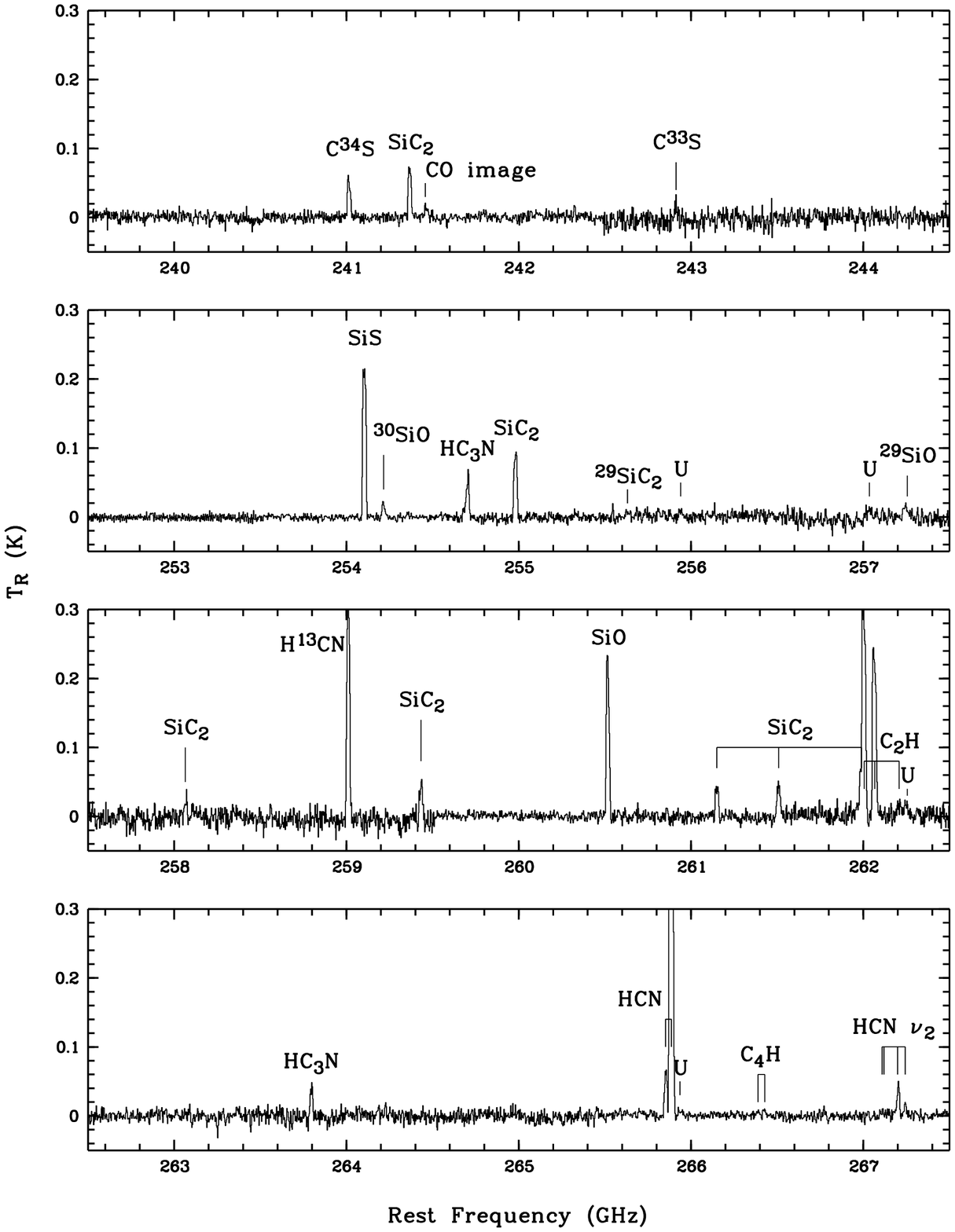,
height=20cm, }
\caption{continued.}
\end{figure*}

\begin{figure*}
\epsfig{file=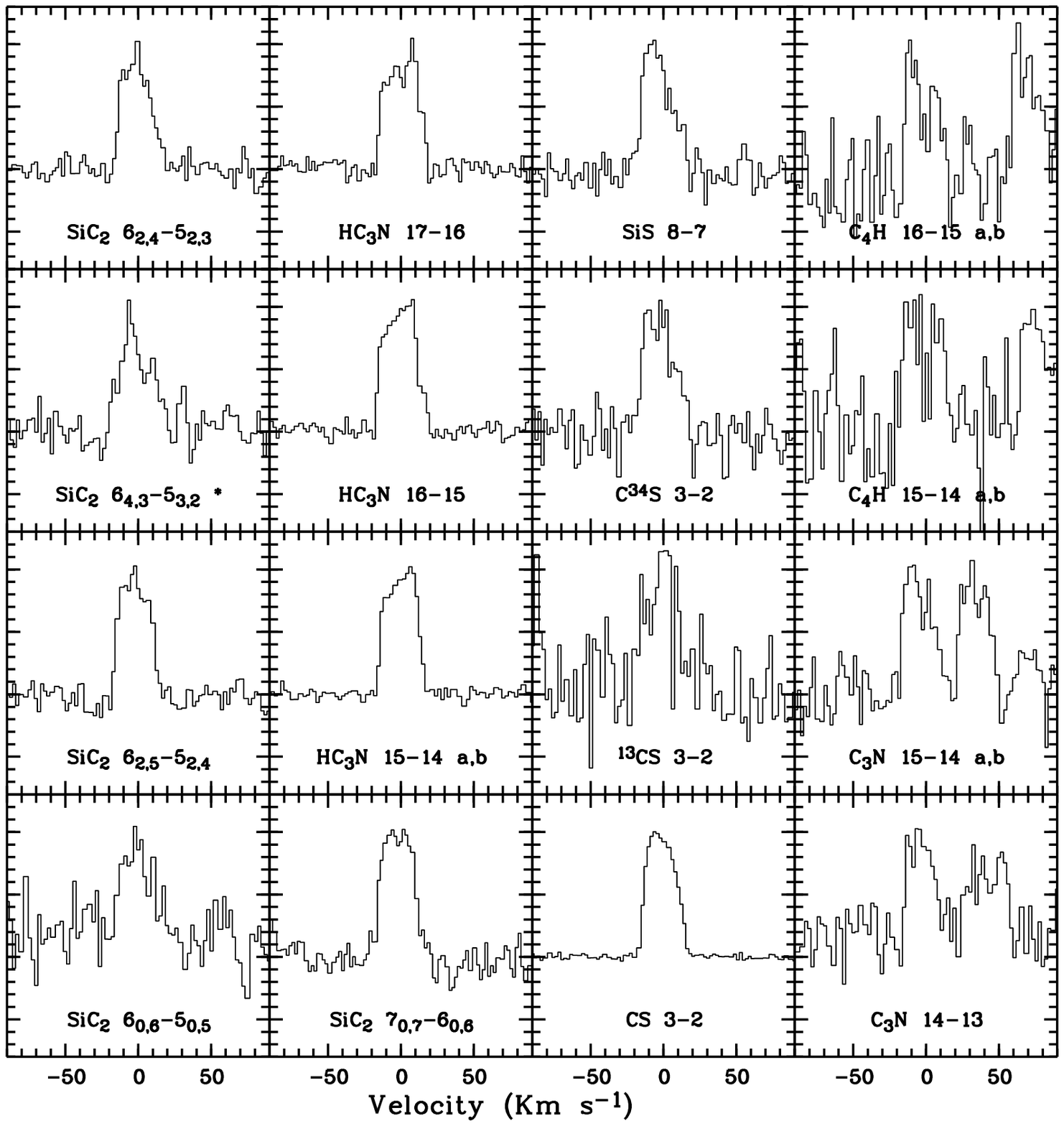,
 height=15cm, }
\caption{Line profiles of the molecular species detected
in the ARO 12\,m spectrum.
The spectral resolution is 1\,MHz. The labeled lines are
shown at zero velocity in each panel.
}
\label{pro_12m}
\end{figure*}

\begin{figure*}
\epsfig{file=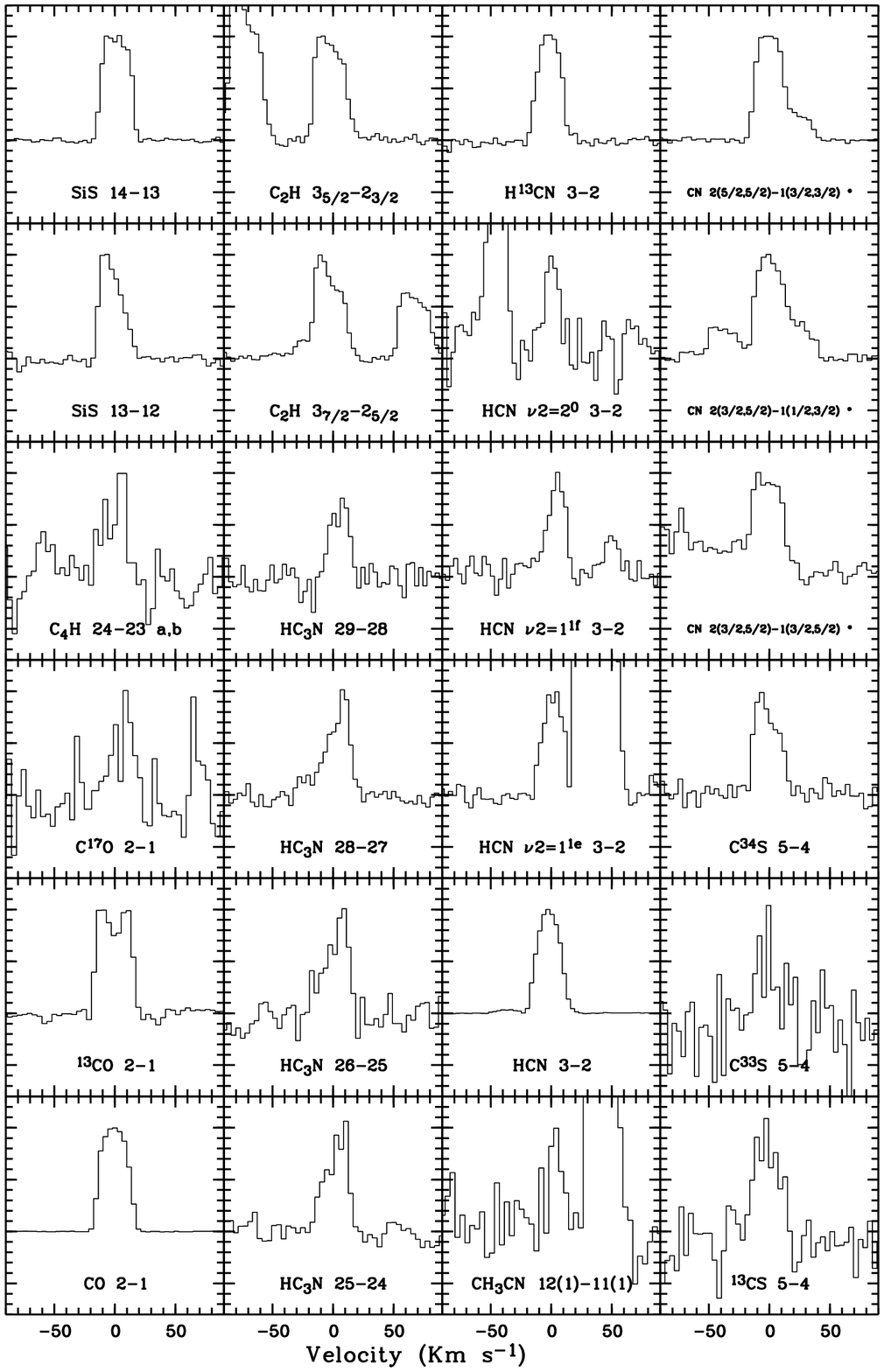,
height=22cm, }
\caption{
Line profiles of the molecular species detected in the SMT spectrum.
The spectral resolution is 3\,MHz. The labeled lines are
shown at zero velocity in each panel.
}
\label{pro_smt}
\end{figure*}

\addtocounter{figure}{-1}
\begin{figure*}
\epsfig{file=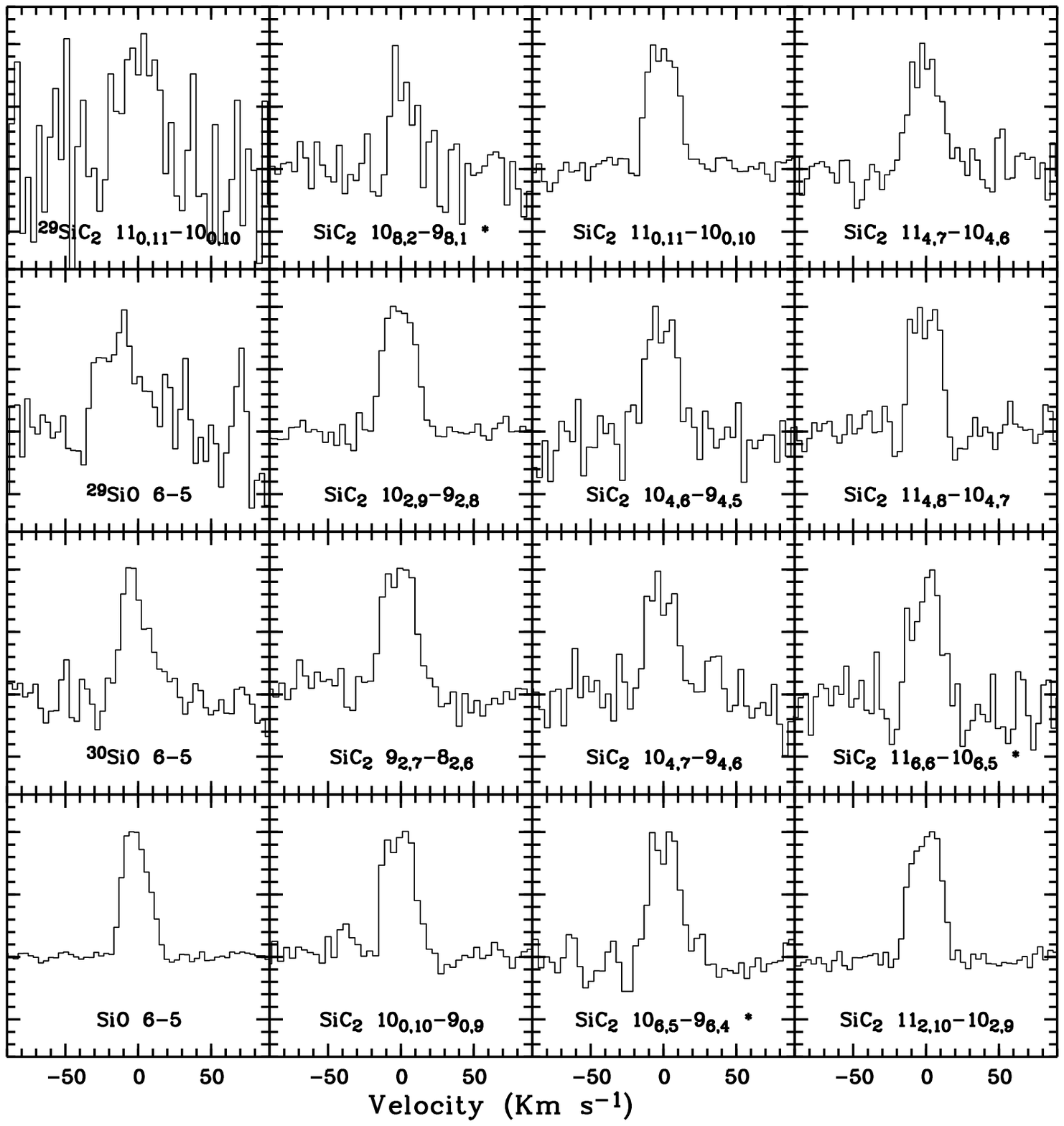,
height=15cm, }
\caption{continued.}
\end{figure*}

\begin{figure*}
\epsfig{file=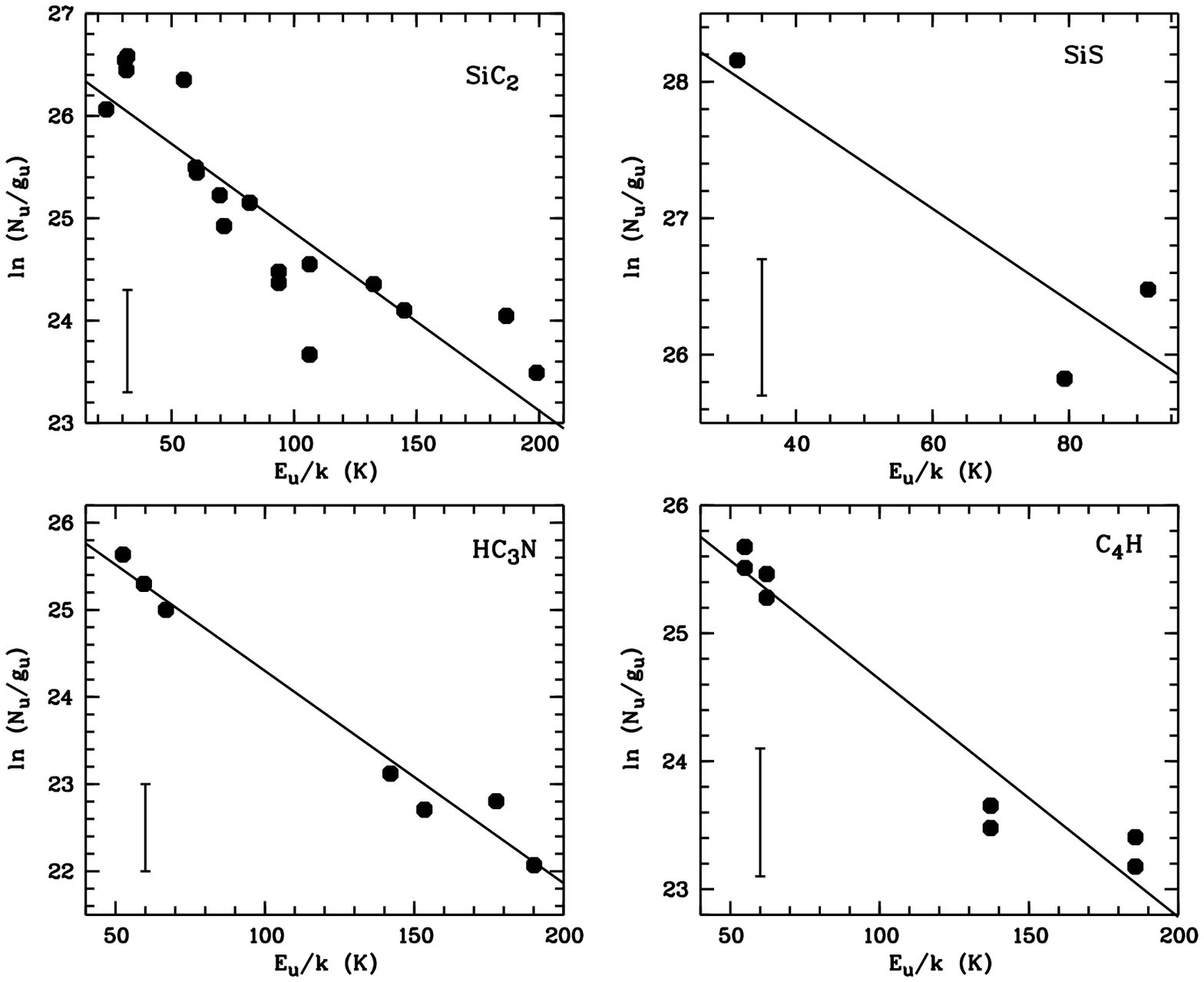, 
height=12cm}
\caption{Rotational diagrams for the detected species in CIT\,6.
Error bars are given  on the lower right.}
\label{dia_cit6}
\end{figure*}

\begin{figure*}
\epsfig{ file=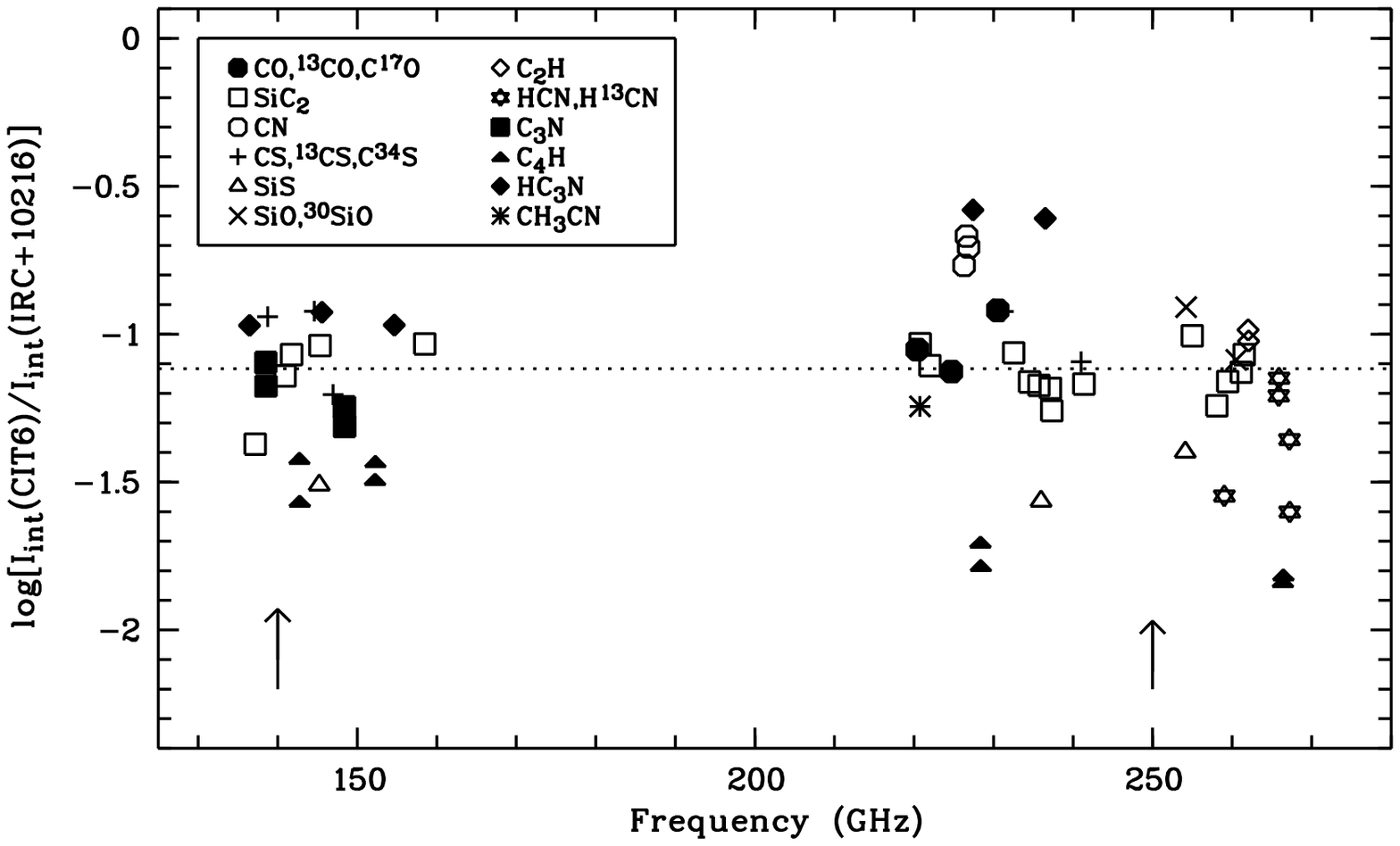,
height=10cm}
\caption{ Integrated intensity ratios of the lines detected
in CIT\,6 and those detected in IRC+10216. The dotted line
represents the average value. A correction of
beam dilution effect will cause the ratios to increase by a factor
of $\xi$.
The $\log\xi$ values for the 12\,m and SMT data are denoted by
the lengths of the arrows in the lower left- and
right-hand corner, respectively.
}
\label{linecomp}
\end{figure*}

\begin{figure*}
\epsfig{ file=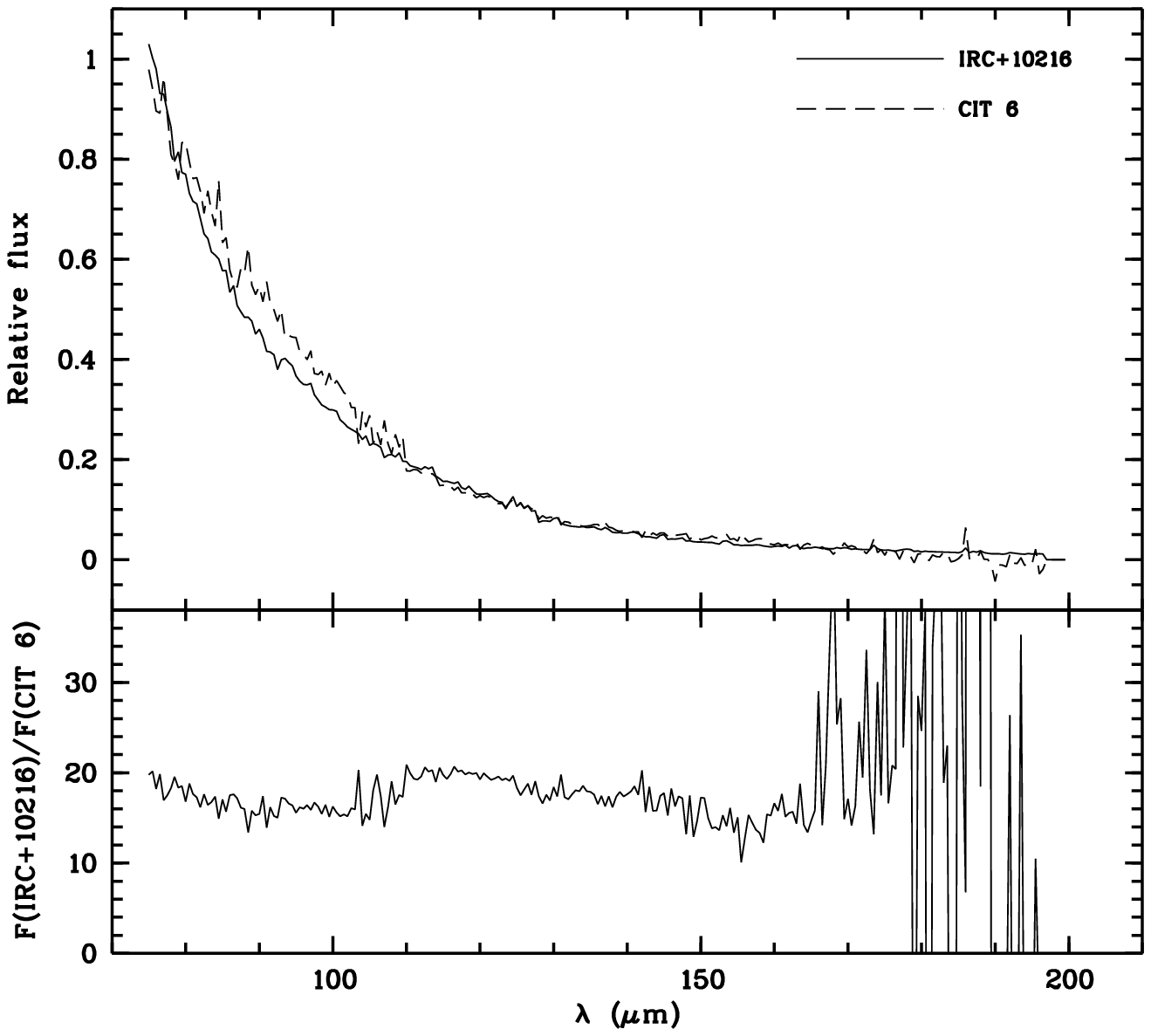, height=12cm}
\caption{{\it Upper panel}: the ISO LWS spectra of IRC+10216 and CIT\,6,
for which the fluxes were normalized such that $F_{76\mu m}=1$;
{\it Lower panel}: the absolute-flux ratio of the two spectra. Note
that the ratio at $\lambda>160$\,$\mu m$ bears large errors.
}
\label{iso_con}
\end{figure*}

\begin{figure*}
\epsfig{ file=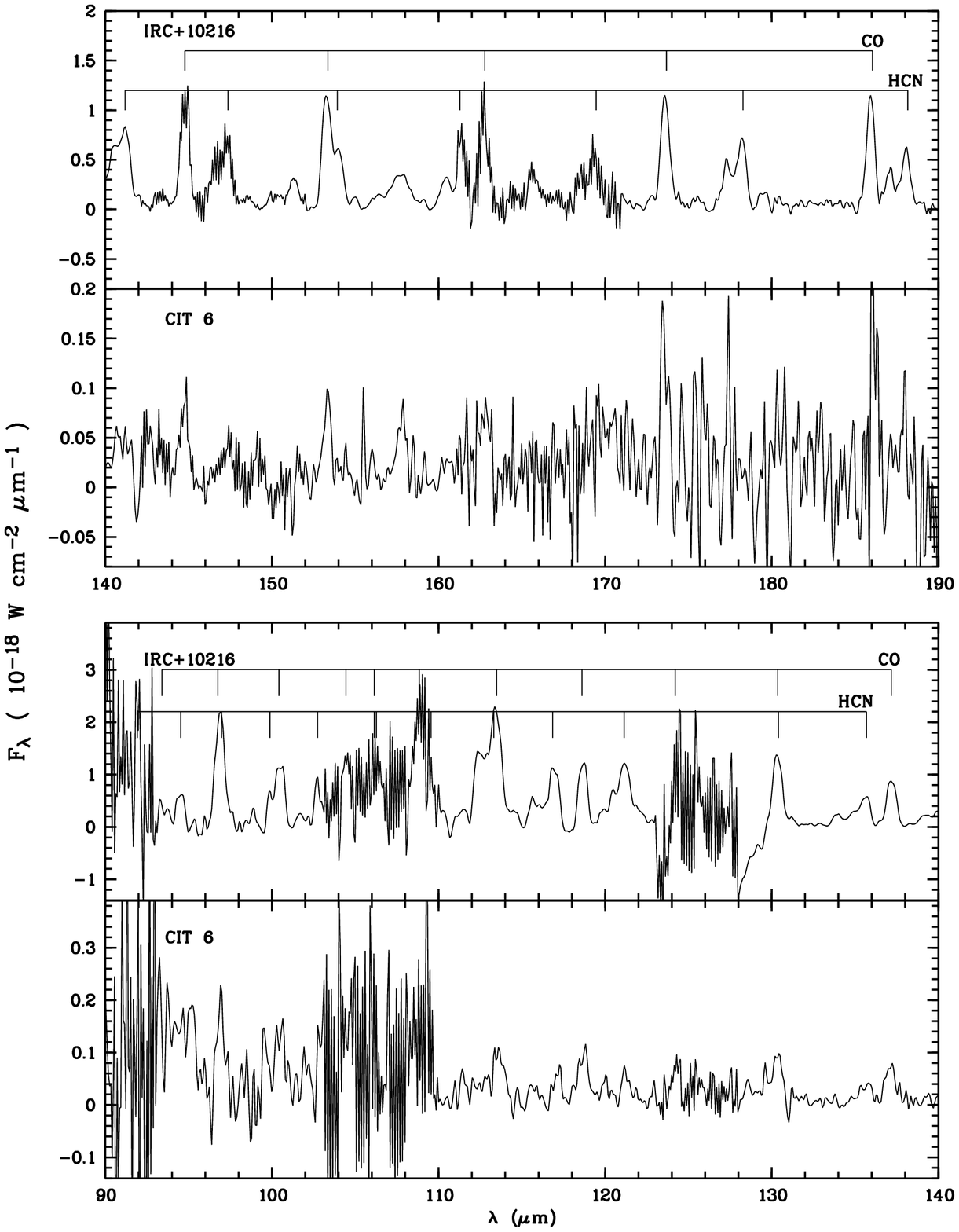,
height=18cm}
\caption{The continuum-subtracted ISO LWS spectra of IRC+10216 and CIT\,6. 
The solid lines mark the CO and HCN emission identified by
\citet{cernicharo96} in the spectrum of IRC+10216.
}
\label{isocomp}
\end{figure*}

\begin{figure*}
\epsfig{file=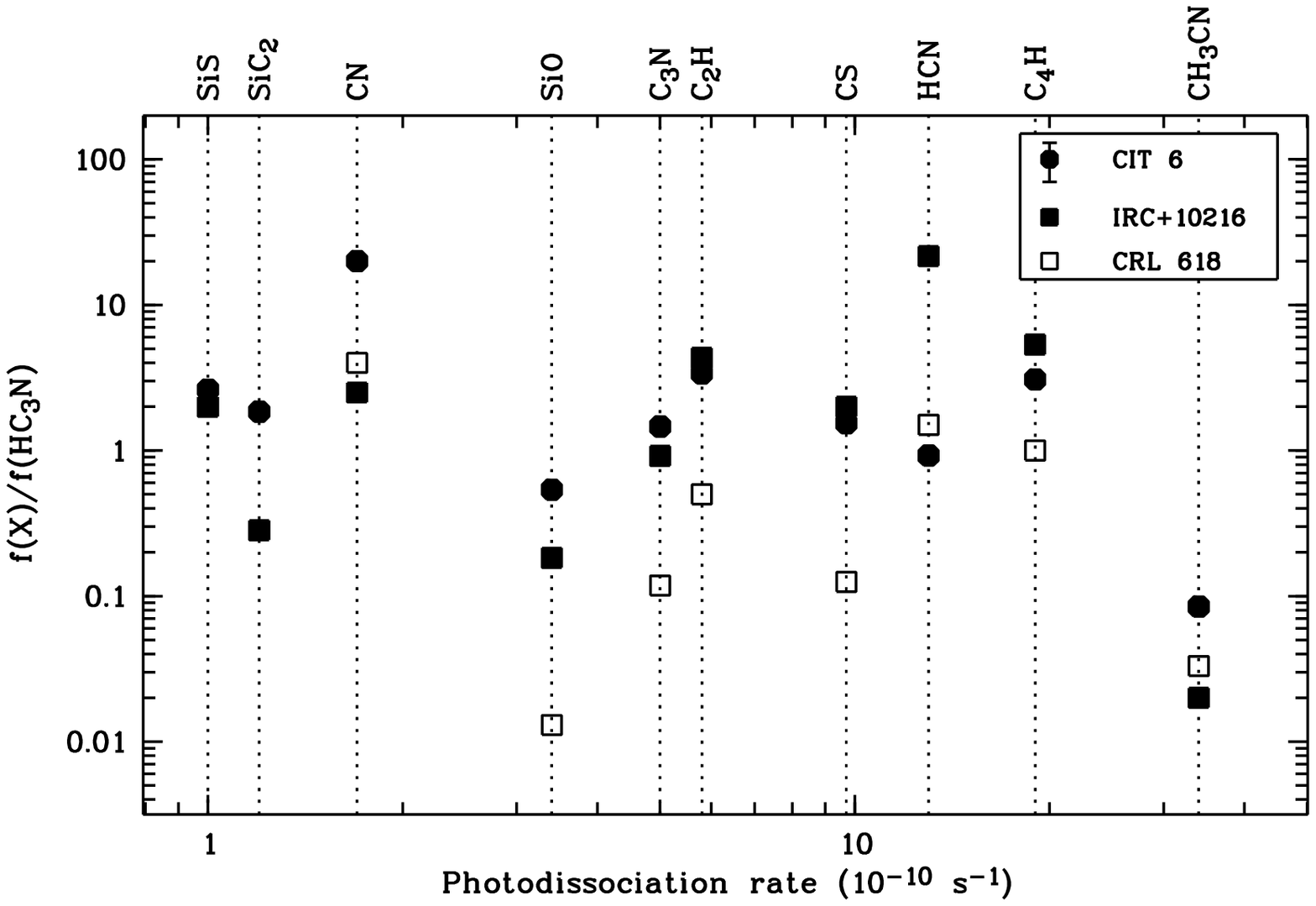, height=10cm}
\caption{Abundances relative to HC$_3$N
versus photodissociation rates at $T=300$\,K. 
For some species, the 
 photodissociation rates have been shifted  a littile bit 
for convenience of display.
The data of IRC+10216 and
CRL\,618 are taken from \citet{woods03} and \citet{pardo07}, respectively.
Typical uncertainty for the data of CIT\,6 is given on the upper right.
Note that the data for SiS and SiC$_2$ in CRL\,618 are lacking.}
\label{compare}
\end{figure*}

\begin{deluxetable}{lllccccc}
\tablecaption{Molecular transitions detected in  CIT\,6.
\label{line}}
\tabletypesize{\scriptsize}
\tablewidth{0pt}
\tablehead{
\colhead{Frequency}&  \colhead{Species} & \colhead{Transition} &  \colhead{$rms$} 
&\colhead{$T_{\rm R}$} &\colhead{$\int T_{\rm R}$d$v$} &  \colhead{$\Delta V_{FWHM}$} &\colhead{Remarks$^a$} 
\\
\colhead{(MHz)} & &\colhead{(upper--lower)} &\colhead{(mK)}  &\colhead{(K)} & \colhead{(K~km/s)}& \colhead{(km/s)} & \\
}
\startdata
230538.0& CO              &   J=2--1                      & 4.2  &  2.555&   67.590 & 23.7\\
220398.7& $^{13}$CO       &   J=2--1                      & 5.6  &  0.163&    4.996 & 29.0\\
224714.4& C$^{17}$O       &   J=2--1                      & 7.1  &  0.022&    0.258 & 25.6\\
137180.8& SiC$_2$         &   6$_{0,6}$--5$_{0,5}$        & 6.8  &  0.040&    1.040 & 24.2\\
140920.1& SiC$_2$         &   6$_{2,5}$--5$_{2,4}$        & 4.8  &  0.062&    1.459 & 22.6\\
141751.5& SiC$_2$         &   6$_{4,3}$--5$_{3,2}$        & 6.1  &  0.073&    1.671 & 23.4 & *\\
141755.4& SiC$_2$         &   6$_{4,2}$--5$_{4,1}$        & ---&  ---  &    ---   & ---& *\\
145325.8& SiC$_2$         &   6$_{2,4}$--5$_{2,3}$        & 6.5 &  0.088&    1.867 & 21.1\\
158499.2& SiC$_2$         &   7$_{0,7}$--6$_{0,6}$        & 9.5 &  0.086&    2.548 & 25.3\\
220773.7& SiC$_2$         &   10$_{0,10}$--9$_{0,9}$      & 5.6&  0.087&    2.282 & 22.8\\
222009.4& SiC$_2$         &   9$_{2,7}$--8$_{2,6}$        & 6.4&  0.069&    1.887& 24.8\\
232534.1& SiC$_2$         &   10$_{2,9}$--9$_{2,8}$       & 4.0&  0.072&    1.937& 25.2\\
234534.0& SiC$_2$         &   10$_{8,2}$--9$_{8,1}$       & 4.8&  0.030&    0.452 & 16.1:& *  \\
234534.0& SiC$_2$         &   10$_{8,3}$--9$_{8,2}$       & ---&  ---  &    ---   & ---& *\\
235713.0& SiC$_2$         &   10$_{6,5}$--9$_{6,4}$       & 5.2&  0.045&    1.101 & 25.2 &*\\
235713.1& SiC$_2$         &   10$_{6,4}$--9$_{6,3}$       & ---&  ---  &    ---   & ---& *\\
237150.0& SiC$_2$         &   10$_{4,7}$--9$_{4,6}$       & 5.8&  0.038&    0.859 & 20.7\\
237331.3& SiC$_2$         &   10$_{4,6}$--9$_{4,5}$       & 5.8&  0.036&    0.771 & 23.5\\
241367.7& SiC$_2$         &   11$_{0,11}$--10$_{0,10}$    & 4.2&  0.074&    1.770 & 22.7\\
254981.5& SiC$_2$         &   11$_{2,10}$--10$_{2,9}$     & 3.9&  0.094&    2.448 & 23.6\\
258065.0& SiC$_2$         &   11$_{8,3}$--10$_{8,2}$      & 9.3&  0.039&    0.471 & 10.5:\\
258065.0& SiC$_2$         &   11$_{8,4}$--10$_{8,3}$      & ---&  ---  &    ---   & --- &*\\
259433.3& SiC$_2$         &   11$_{6,6}$--10$_{6,5}$      & 7.8&  0.054&    1.301 & 27.5 &*\\
259433.3& SiC$_2$         &   11$_{6,5}$--10$_{6,4}$      & ---&  ---  &    ---   & --- &*\\
261150.7& SiC$_2$         &   11$_{4,8}$--10$_{4,7}$      & 4.1&  0.044&    1.049 & 21.5\\
261509.3& SiC$_2$         &   11$_{4,7}$--10$_{4,6}$      & 4.1&  0.051&    1.270 & 26.5\\
261990.7& SiC$_2$         &   12$_{0,12}$--11$_{0,11}$    & 7.1&  0.387&    9.945 & 26.1 &\$\\
238347.0& $^{29}$SiC$_2$ &   11$_{0,11}$--10$_{0,10}$     & 5.3&  0.013 &   0.366  & 35.3\\
255631.6& $^{29}$SiC$_2$ &   11$_{6,6}$--10$_{6,5}$       & 4.2&  0.010 &   0.127  & 32.9:&* \\
255631.8& $^{29}$SiC$_2$ &   11$_{6,5}$--10$_{6,4}$        & ---&  ---  &    ---   & --- &*\\
226287.4& CN              &   N(J,F)=2(3/2,1/2)--1(3/2,1/2)&4.9 &  0.034&    1.278 & 53.0: &*\\
226298.9& CN              &   N(J,F)=2(3/2,1/2)--1(3/2,3/2)  & ---&  ---  &    ---   & --- &*\\
226303.0& CN              &   N(J,F)=2(3/2,3/2)--1(3/2,1/2)  & ---&  ---  &    ---   & --- &*\\
226314.5& CN              &   N(J,F)=2(3/2,3/2)--1(3/2,3/2)  & ---&  ---  &    ---   & --- &*\\
226332.5& CN              &   N(J,F)=2(3/2,3/2)--1(3/2,5/2) &4.9 &  0.052&    1.819 & 25.6 & *\\
226341.9& CN              &   N(J,F)=2(3/2,5/2)--1(3/2,3/2)   & ---&  ---  &    ---   & --- &*\\
226359.9& CN              &   N(J,F)=2(3/2,5/2)--1(3/2,5/2)    & ---&  ---  &    ---   & --- &*\\
226616.6& CN              &   N(J,F)=2(3/2,1/2)--1(1/2,3/2)  &4.9 &  0.053&    1.593 & 26.8 &*\\
226632.2& CN              &   N(J,F)=2(3/2,3/2)--1(1/2,3/2)    & ---&  ---  &    ---   & --- &*\\
226659.6& CN              &   N(J,F)=2(3/2,5/2)--1(1/2,3/2)  &4.9 &  0.178&    6.092 & 29.3 & * \\
226663.7& CN              &   N(J,F)=2(3/2,1/2)--1(1/2,1/2)     & ---&  ---  &    ---   & --- &*\\
226679.3& CN              &   N(J,F)=2(3/2,3/2)--1(1/2,1/2)     & ---&  ---  &    ---   & --- &*\\
226874.2& CN              &   N(J,F)=2(5/2,5/2)--1(3/2,3/2)  &4.9 &  0.394&   12.176 & 24.6& *\\
226874.8& CN              &   N(J,F)=2(5/2,7/2)--1(3/2,5/2)     & ---&  ---  &    ---   & --- &*\\
226875.9& CN              &   N(J,F)=2(5/2,3/2)--1(3/2,1/2)     & ---&  ---  &    ---   & --- &*\\
226887.4& CN              &   N(J,F)=2(5/2,3/2)--1(3/2,3/2)     & ---&  ---  &    ---   & --- &*\\
226892.1& CN              &   N(J,F)=2(5/2,5/2)--1(3/2,5/2)     & ---&  ---  &    ---   & --- &*\\
226905.4& CN              &   N(J,F)=2(5/2,3/2)--1(3/2,5/2)    & ---&  ---  &    ---   & --- &*\\
146969.0& CS              &   J=3--2                      & 4.8  &  0.303&    7.058 & 21.6\\
138739.3& $^{13}$CS       &   J=3--2                      & 4.6  &  0.023&    0.495 & 21.0\\
231221.0& $^{13}$CS       &   J=5--4                      & 5.0  &  0.027&    0.846 & 27.6\\
242913.6& C$^{33}$S       &   J=5--4                      & 7.4  &  0.032&    0.563:& 12.4:\\
144617.1& C$^{34}$S       &   J=3--2                      & 4.9 &  0.040&    1.039 &  26.1\\
241016.1& C$^{34}$S       &   J=5--4                      & 4.6&  0.062&    1.315 & 23.1\\
145227.0& SiS             &   J=8--7                      & 6.5&  0.067&    1.745 & 22.1\\
235961.1& SiS             &   J=13--12                    & 5.2&  0.119&    2.598 & 20.6\\
254102.9& SiS             &   J=14--13                    & 3.6&  0.213&    5.791 & 25.0\\
260518.0& SiO             &   J=6--5                      & 4.8&  0.233&    5.233 & 20.3\\
254216.7& $^{30}$SiO      &   J=6--5                      & 3.6&  0.022&    0.581 & 21.5\\
257255.2& $^{29}$SiO      &   J=6--5                      & 6.9&  0.021&    0.583 & 30.3\\
262005.3& C$_2$H          &   N$_{\rm J}$=3$_{7/2}$--2$_{5/2}$ & 7.1&  0.387&    9.945 & 26.1 &\$\\
262066.1& C$_2$H          &   N$_{\rm J}$=3$_{5/2}$--2$_{3/2}$ & 7.1&  0.244&    6.596 & 25.4\\
262208.4& C$_2$H          &   N$_{\rm J}$=3$_{5/2}$--2$_{5/2}$ & 6.0&  0.024&    0.539:& 22.9:\\
265886.4& HCN             &   J=3--2                      & 3.7&  1.959&   43.942 & 21.2 \\
265852.8& HCN             &   $\nu_2$=1$^{1e}$ J=3--2     & 3.7&  0.067&    1.392 & 20.4\\
267109.1& HCN             &   $\nu_2$=2$^{2f}$ J=3--2     & 4.2 &  0.011&    0.098:&  22.5:& *\\
267120.1& HCN             &   $\nu_2$=2$^{2e}$ J=3--2     & ---&  ---  &    ---   & --- &*\\
267199.3& HCN             &   $\nu_2$=1$^{1f}$ J=3--2     & 4.2 &  0.050&    0.876 &14.4\\
267243.2& HCN             &   $\nu_2$=2$^0$ J=3--2        & 4.2 &  0.020&    0.318 &14.0\\
259011.8& H$^{13}$CN      &   J=3--2                      & 9.6 &  0.296&    7.120 & 20.8\\
138515.7& C$_3$N          &   N=14--13 a                  & 7.3 &  0.039&    0.747 & 21.5\\
138534.5& C$_3$N          &   N=14--13 b                  & 7.3 &  0.031&    0.674 & 25.5\\
148409.1& C$_3$N          &   N=15--14 a                  & 4.2 &  0.021&    0.537 & 29.9\\
148427.8& C$_3$N          &   N=15--14 b                  & 4.2 &  0.021&    0.521 & 35.6\\
142728.8& C$_4$H          &   N=15--14 a                  & 6.0 &  0.022&    0.567 & 33.7\\
142767.3& C$_4$H          &   N=15--14 b                  & 6.0 &  0.021&    0.450 & 23.1\\
152243.6& C$_4$H          &   N=16--15 a                  & 5.7 &  0.018&    0.433 & 23.6\\
152282.1& C$_4$H          &   N=16--15 b                  & 5.7 &  0.020&    0.505 & 27.2\\
228348.6& C$_4$H          &   N=24--23 a                  & 4.1&  0.015&    0.224:& 19.7:\\
228387.0& C$_4$H          &   N=24--23 b                  & 4.1&  0.035&    0.256:& 13.1:\\
266389.9& C$_4$H          &   N=28--27 a                  & 3.9&  0.008&    0.140:&29.9: \\
266428.2& C$_4$H          &   N=28--27 b                  & 3.9&  0.009&    0.175:& 19.1:\\
136464.4& HC$_3$N         &   J=15--14                    & 6.6&  0.160&    4.131 & 24.0\\
145560.9& HC$_3$N         &   J=16--15                     &6.5&  0.139&    3.772 & 24.1\\
154657.3& HC$_3$N         &   J=17--16                     & 7.0&  0.131&    3.165 & 24.4\\
227418.9& HC$_3$N         &   J=25--24                    & 6.3&  0.064&    1.404 & 22.0\\
236512.8& HC$_3$N         &   J=26--25                    & 6.2 &  0.053&    1.078 & 21.7\\
254699.5& HC$_3$N         &   J=28--27                     &3.9&  0.068&    1.160 & 21.7\\
263792.3& HC$_3$N         &   J=29--28                    & 7.9 &  0.048&    0.990 & 19.6\\
220743.0& CH$_3$CN        &12(1)--11(1)                   & 5.6 &  0.023&    0.290 & 17.6:& *\\
220747.2& CH$_3$CN        &12(0)--11(0)                   & --  &  --   &    --    & -- & *\\
\enddata
\tablenotetext{{\it a}}{ $*$--unsolved hyperfine structure lines; \$--blended
with other species.}
\end{deluxetable}

\begin{deluxetable}{lllll}
\tablecaption{Unidentified lines.
\label{uline}}
\tabletypesize{\footnotesize}
\tablewidth{0pt}
\tablehead{
\colhead{Frequency} &   \colhead{$rms$} &\colhead{$T_{\rm R}$} &\colhead{$\int T_{\rm R}$d$v$} & \colhead{$\Delta V_{FWHM}$} \\
\colhead{(MHz)} &  \colhead{(mK)}      &\colhead{(K)}         & \colhead{(K~km/s)}  & \colhead{(km/s)}\\
}
\startdata
148444  &    4.5    &  0.008&    0.249 & 14.8\\
257035  &    7.1    &  0.015&    0.442  &38.6 \\
255940  &    4.9    &  0.012&    0.207  &29.4 \\
262255  &    6.0    &  0.022&    0.501 & 26.7\\
265936  &    3.7    &  0.012&   0.214   & 37.3 \\
\enddata
\end{deluxetable}

\begin{deluxetable}{lccrccr}
\tablecaption{Excitation temperatures, column densities
and abundances with respect to H$_2$$^a$.
\label{col_cit6}}
\tabletypesize{\footnotesize}
\tablewidth{0pt}
\tablehead{
\colhead{Species} & \colhead{$T_{\rm ex}$\,(K)$^b$} & \multicolumn{2}{c}{$N$\,(cm$^{-2}$)}&
& \multicolumn{2}{c}{$f_{\rm X}$}\\
\cline{3-4} \cline{6-7}
& &\colhead{This paper$^c$} & \colhead{F94$^d$} & &\colhead{This paper$^c$} & \colhead{W03$^d$}\\
}
\startdata
SiC$_2$    & 57.6    &  1.53(14) & \nodata         & &2.4(-6)  & 3.1(-6)\\
$^{29}$SiC$_2$    &  \nodata&  2.26(13) &  \nodata & &3.6(-7)    &\nodata \\
SiS        & 29.6    &  2.96(14) & \nodata         & &3.4(-6)  & 3.1(-6)\\
C$_4$H     & 53.9    &  3.04(14) & $<5.54$(13)     & &4.0(-6)  & $<1.7$(-6)\\
HC$_3$N    & 41.0    &  7.69(13) & $2.7\pm2.3$(14) & &1.3(-6)  & 2.4(-6)\\
CO         & \nodata&  1.63(17)  & \nodata         & &\nodata  & \nodata \\
$^{13}$CO  & \nodata&  1.35(16)  & \nodata         & &\nodata  & \nodata  \\
C$^{17}$O  & \nodata&  6.89(14)  & \nodata         & &\nodata  & \nodata  \\
CS         &  \nodata&  1.72(14) &  \nodata        & &2.0(-6)  & 2.5(-6)\\
$^{13}$CS  &  \nodata&  1.28(13) &  \nodata        & &1.5(-7)  & \nodata\\
C$^{33}$S  &  \nodata&  6.18(12):&  \nodata        & &7.2(-8):  & \nodata\\
C$^{34}$S  &  \nodata&  2.58(13) &  \nodata        & &3.0(-7)  & \nodata\\
SiO        &  \nodata&  2.15(13) &  \nodata        & &7.0(-7)  & 1.0(-6)\\
$^{30}$SiO &  \nodata&  2.45(12) &  \nodata        & &8.0(-8)  & \nodata\\
$^{29}$SiO &  \nodata&  2.43(12) &  \nodata        & &7.9(-8)  & \nodata\\
HCN        &  \nodata&  1.10(14) & 5.13e15         & &1.2(-6)  & 1.1(-5)\\
H$^{13}$CN &  \nodata&  1.83(13) & \nodata         & &2.0(-7)  & 3.0(-7)\\
CN         &  \nodata&  1.22(15) & \nodata         & & 2.6(-5) & 2.0(-5)\\
C$_2$H     &  \nodata&  7.43(14) & $4.0\pm2.0$(14) & & 5.4(-6) & 6.9(-6)\\
C$_3$N     &  \nodata&  2.16(13) & \nodata         & & 1.9(-6) & 2.6(-6)\\
CH$_3$CN   &  \nodata&  7.88(12) & \nodata         & & 1.1(-7) & $<1.3$(-7)\\
\enddata                           
\tablenotetext{{\it a}}{$x(y)$ represents $x\times10^y$;}
\tablenotetext{{\it b}}{A constant excitation temperature of 40\,K was
assumed for the species for which 
the rotation diagrams cannot be obtained; }
\tablenotetext{{\it c}}{For the species  with optically thick 
emission (e.g. CO and HCN),
this gives the lower limits;}
\tablenotetext{{\it d}}{F94: from Fukasaku et al. (1994); W03: from Woods et al. (2003).}
\end{deluxetable}

\begin{deluxetable}{lllll}
\tablecaption{Isotopic abundance ratios.
\label{isoto_cit6}}
\tablewidth{0pt}
\tablehead{
\colhead{Isotopic ratio} & \multicolumn{2}{c}{CIT\,6} & \colhead{IRC+10216$^a$} & \colhead{Solar$^b$}\\
\cline{2-3}
& \colhead{Sepcies}& \colhead{Value$^c$}\\
}
\startdata
$^{12}$C/$^{13}$C & $^{12}$C$^{34}$S/$^{13}$C$^{32}$S  &45.4$\pm4.9^d$   & 45$\pm3$       & 89   \\
 &$^{12}$CO/$^{13}$CO  & 12.1$\pm1.3$$^e$  & \nodata       & \nodata \\
 &$^{12}$CS/$^{13}$CS  & 13.4$\pm1.7$$^e$   &  \nodata      & \nodata \\
 &H$^{12}$CN/H$^{13}$CN& 6.0$\pm0.7$$^e$   & \nodata       & \nodata \\
$^{16}$O/$^{17}$O &  $^{13}$C$^{16}$O/$^{12}$C$^{17}$O   & 890$\pm97^f$  &  \nodata       & 2680 \\
                  &  C$^{16}$O/C$^{17}$O   & 237 $\pm26^e$  &  \nodata       & \nodata \\
$^{29}$Si/$^{30}$Si & $^{29}$SiO/$^{30}$SiO & 1.0$\pm0.4$  & 1.45$\pm0.13$  & 1.52 \\
$^{28}$Si/$^{30}$Si & $^{28}$SiO/$^{30}$SiO & 8.8$\pm1.9$$^e$   & 20.3$\pm2.0^e$   & 29.9 \\
$^{28}$Si/$^{29}$Si &  $^{28}$SiO/$^{29}$SiO&   8.9$\pm2.8^e$ & 15.4$\pm2.0^e$ & 19.6 \\
                &  $^{28}$SiC$_2$/$^{29}$SiC$_2$& 6.7$\pm3.1^e$   &  \nodata & \nodata  \\
$^{32}$S/$^{34}$S  & C$^{32}$S/C$^{34}$S  & 6.7$\pm1.2$$^e$   & 21.8$\pm2.6$   & 22.5 \\
$^{33}$S/$^{34}$S  & C$^{33}$S/C$^{34}$S  & 0.2$\pm0.2$   & 0.18$\pm0.1$   & 0.18 \\
\enddata
\tablenotetext{{\it a}}{From Cernicharo et al. (2000);}
\tablenotetext{{\it b}}{From Lodders (2003);}
\tablenotetext{{\it c}}{The errors are estimated based on the measurement 
and calibration uncertainties;}
\tablenotetext{{\it d}}{Assume that the $^{34}$S/$^{32}$S ratio is solar;}
\tablenotetext{{\it e}}{Should be treated as lower limits due to opacity effect;}
\tablenotetext{{\it f}}{Adopted: $^{12}$C/$^{13}$C=45.4.}
\end{deluxetable}


\begin{thebibliography}{}

\bibitem[Ag\'undez \& Cernicharo(2006)]{agundez06} 
Ag\'undez, M., \& Cernicharo, J. 2006, \apj, 650, 374

\bibitem[Alksnis(1995)]{alksnis95} Alksnis, A. 1995, Baltic Astron., 4, 79

\bibitem[Bachiller et al.(1997)]{bachiller97} Bachiller, R., Fente, A.,
Bujarrabal, V., Colomer, F., Loup, C., Omont, A., \& de Jong, T.
1997, \aaps,  319, 235


\bibitem[Balser et al.(2002)]{balser02} Balser, D. S., McMullin, J. P., \& Wilson, T. L. 2002, \apj, 572, 326

\bibitem[Bieging et al.(2000)]{bieging00} Bieging J. H., Shaked, S.,
\& Gensheimer, P. D. 2000, \apj, 543, 897

\bibitem[Boothroyd \& Sackmann(1999)]{boothroyd99} Boothroyd, A. I., \& Sackmann, I.-J. 1999, \apj, 510, 232
	
\bibitem[Bujarrabal et al.(1994)]{bujarrabal94} Bujarrabal, V., Fuente, A., 
\& Omont, A. 1994, \aap, 285, 247

\bibitem[Busso(2006)]{busso06} Busso, M. M.  2006,
in IAU Symp. 234 Planetary Nebulae, eds.  M. J. Barlow,
\& R. H. M\'{e}ndez (Cambridge: Cambridge University Press), P.91

\bibitem[Cernicharo et al.(1996)]{cernicharo96}
Cernicharo, J., Barlow, M. J., Gonz\'alez-Alfonso, E. et al.
1996, \aap, 315, L201

\bibitem[Cernicharo et al.(2000)]{cernicharo00}
Cernicharo, J., Gu{\' e}lin, M., \& Kahane, C. 2000, \aaps, 142, 181

\bibitem[Charbonnel(1995)]{charbonnel95}
Charbonnel, C. 1995, \apj, 453, L41

\bibitem[Charbonnel \& do Nascimento(1998)]{charbonnel98}
Charbonnel, C., \& do Nascimento, J. D. 1998, \aap, 336, 915


\bibitem[Cherchneff(2006)]{cherchneff06} 
Cherchneff, I. 2006, \aap, 456, 1001

	
\bibitem[Cohen \& Hitchon(1996)]{cohen96}
Cohen, M., \& Hitchon, K. 1996, \aj, 111, 962


\bibitem[Dayal \& Bieging(1993)]{dayal93} Dayal, A., \& Bieging,
J. H. 1993, \apj, 407, L37

\bibitem[Dyck et al.(1971)]{dyck71} Dyck, H. M., Forbes, F. F., \& Shawl, S. J. 1971, AJ, 76, 901

\bibitem[Frost et al.(1998)]{frost98} Frost, C. A., Cannon, R. C., Lattanzio, J. C., Wood, P. R., \& Forestini, M. 1998, \aap, 332, L17

\bibitem[Fukasaku et al.(1994)]{fukasaku94} Fukasaku, S.,
Hirahara, Y., Masuda, A. et al. 1994, \apj, 437, 410

\bibitem[Glassgold(1996)]{glassgold96}
Glassgold, A. E. 1996, \araa, 34, 241

\bibitem[Gonz\'alez Delgado et al.(2003)]{gonzalez03}
Gonz\'alez Delgado, D., Olofsson, H., Kerschbaum, F.,
Sch\"oier, Lindqvist, M., \& Groenewegen, M. A. T.
2003, \aap, 411, 123

\bibitem[Groenewegen et al.(1996)]{groenewegen96}
Groenewegen, M. A. T., Baas, F., de JONG, T., \& Loup, C. 
1996, \aap, 306, 241

\bibitem[Groenewegen et al.(1998)]{groen98}
Groenewegen, M.A.T., Whitelock, P.A., Smith, C.H., \&  Kerschbaum, F. 1998, MNRAS, 293, 18



\bibitem[He et al.(2008)]{he08} He, J.-H., 
Dinh-V-Trung, Kwok, S., M\"uller, H. S. P., Zhang, Y., Hasegawa, T., Peng,
T. C., \& Huang, Y. C. 2008, \apjs, 177, 275


\bibitem[Henkel et al.(1985)]{henkel85} Henkel, C., Matthews, H. E.,
Morris, M., Terebey, S., \& Fich, M. 1985 \aap, 147, 143

\bibitem[Herpin et al.(2002)]{herpin02}
Herpin, F., Goicoechea, J. R., \& Cernicharo, J. 2002, \apj, 577, 961

\bibitem[Josselin \& Bachiller(2003)]{josselin03} Josselin, E., \&
Bachiller, R. 2003, \aap, 397, 659

\bibitem[Kahane et al.(1992)]{kahane92} Kahane, C., Cernicharo, J.,
G\'omez-Gonz\'alez, J., \& Gu\'elin, M. 1992, A\&AS, 256, 235


\bibitem[Keene et al.(1998)]{keene98}
Keene, J., Schilke, P., Kooi, J., Lis, D. C.,
 Mehringer, D. M., \& Phillips, T. G. 1998, \apj, 494, L107

\bibitem[Kruszewski(1968)]{kruszewski68} Kruszewski, A. 1968, PASP, 80, 560

\bibitem[Kwok(2004)]{kwok04} Kwok, S. 2004, Nature, 430, 985

\bibitem[Lagadec et al.(2005)]{lagadec05}
Lagadec, E., M{\' e}karnia, D.,  de Freitas Pacheco, J. A., \&  Dougados, C,
2005, A\&A, 433, 553 

\bibitem[Lindqvist et al.(2000)]{lin00}
Lindqvist, M., Sch\"oier, F.L., Lucas, R., \&  Olofsson, H. 2000, \aap, 361, 1036

\bibitem[Lodders(2003)]{lodders} Lodders, K. 2003, \apj, 591, 1220

\bibitem[Lucas \& Gu\'elin(1999)]{gue99}
Lucas, R., Gu\'elin, M. 1999, in IAU Symp. 191 Asymptotic Giant Brnach Stars, T. Le Bertre,  A. L\'ebre, C. Waelkens (eds), ASP, p. 305

\bibitem[MacKay \& Charnley(1999)]{mackay99}
MacKay, D. D. S., \& Charnley, S. B. 1999, \mnras, 302, 793

\bibitem[Monnier et al.(2000)]{monnier00}
Monnier, J. D., Tuthill, P. G., \& Danchi, W. C. 2000, \apj, 545, 957

\bibitem[M\"uller et al.(2001)]{muller01} M\"uller, H. S. P., Thorwirth, S., Roth, D. A., \& Winnewisser, G.,
2001, \aap, 370, L49

\bibitem[M\"uller et al.(2005)]{muller05} M\"uller, H. S. P., Schl\"oder, F., Stutzki, J.,  \& Winnewisser, G., 2005, J. Mol. Struct. 742, 215

\bibitem[Nejad \& Millar(1987)]{nejad87} Nejad, L. A. M.,
\& Millar, T. J. 1987, \aap, 183, 279


\bibitem[Nummelin et al.(1998)]{num98}
Nummelin, A., Bergman, P., Hjalmarson, A. et al. 1998, \apjs, 117, 427

\bibitem[Olofsson(1996)]{olofsson96}
Olofsson, H. 1996, in IAU Symp. 178, Molecules in Astrophysics: Probes \& Processes, ed. E. van Dishoeck (Dordrecht: Kluwer), 457

\bibitem[Olofsson(1997)]{olo97}
Oloffson, H. 1997, Astrophys. Sp. Sci., 251, 31


\bibitem[Pardo et al.(2007)]{pardo07} Pardo, J. R., Cernicharo, J.,
Goicoechea, J. R., Gu{'e}lin, M., \& Ramos, A. A. 2007, \apj,  661, 250


\bibitem[Pickett et al.(1998)]{pickeet98} Pickett, H. M., Poynter, R. L., Cohen, E. A., Delitsky, M. L.,  Pearson, J. C., \& Muller, H. S. P., 
1998, J. Quant. Spectrosc. \& Rad. Transfer, 60, 883

\bibitem[Sackmann \& Boothroyd(1999)]{sackmann99} Sackmann, I.-J., \& Boothroyd, A. I. 1999, ApJ, 510, 217


\bibitem[Sch\"oier et al.(2006)]{schoier06}
Sch\"oier, F. L.,  Olofsson, H., \& Lundgren, A. A.
2006, \aap, 247, 255

\bibitem[Sch\"oier et al.(2002)]{schoier02}
Sch\"oier, F. L.,  Ryde, N., \& Olofsson, H., 2002, \aap, 391, 577

\bibitem[Sch\"oier et al.(2007)]{schoier07}
Sch\"oier, F. L. , Bast, J., Olofsson, H., \& Lindqvist, M.
2007, \aap, 473, 871

\bibitem[Schmidt et al.(2002)]{schmidt02}
Schmidt, G. D., Hines, D. C., \& Swift, S. 2002, \apj, 576, 429

\bibitem[Sopka et al.(1989)]{sopka89} 
Sopka, R. J., Olofsson, H., Johansson, L. E. B., Nguyen-Q-Rieu, \&  Zuckerman, B. 1989, \aap, 210, 78


\bibitem[Teyssier et al.(2006)]{teyssier06} 
Teyssier, D, , Hernandez, R., Bujarrabal, V., Yoshida, H., \& Phillips,
T. G. 2006, \aap, 450, 167

\bibitem[Ulrich et al.(1966)]{ulrich66}
Ulrich, B.T., Neugebauer, G., McCammono, D., Leighton, R. B., Hughes, E. E., \&  Becklin, E. 1966, \apj, 146, 288

\bibitem[Wannier \& Sahai(1987)]{wannier87} Wannier, P. G., \&
Sahai, R. 1987, \apj, 319, 367

\bibitem[Willacy \& Cherchneff(1998)]{willacy98} Willacy, K., \&
Cherchneff, I. 1998, \aap, 330, 676

\bibitem[Winters et al.(2002)]{winters02} Winters, J. M.,  Le Bertre, T., 
Nyman, L.-{\AA}, Omont, A., \& Jeong, K. S. 2002, \aap, 388, 609

\bibitem[Woods et al.(2003)]{woods03} Woods, P. M., Sch\"oier, F. L.,
Nyman, L.-{\AA}, \& Olofsson, H. 2003, \aap, 402, 617

\bibitem[Wootten et al.(1980)]{wootten80} Wootten, A., Bozyan, E. P.,
\& Garrett, D. B. 1980, ApJ, 239, 844 

\bibitem[Ziurys et al.(2007)]{ziu07}
Ziurys, L. M., Milam, S. N., Apponi, A. J., \&  Woolf, N. J. 2007, Nature, 447, 1094


\end{thebibliography}
\end{document}